\documentclass[aap,MSNbibl,seceqn,citesort,dvips]{arximspdf}
\usepackage{subeqn}
\usepackage{graphicx}

\doi{10.1214/12-AAP857} 
\volume{23}
\issue{3}
\pubyear{2013}
\firstpage{859}
\lastpage{894}

\makeatletter
\newcommand{\eqref}[1]{(\ref{#1})}
\newtheorem{thmm}{Theorem}[section]
\newtheorem{lem}[thmm]{Lemma}
\newtheorem{prop}[thmm]{Proposition}
\newproclaim{assumption}[thmm]{Assumption}
\newproclaim{defn}[thmm]{Definition}
\newproclaim{rmk}[thmm]{Remark}
\newproclaim{example}[thmm]{Example}

\newcommand{\rarr}{\rightarrow}
\newcommand{\E}{\mathbb{E}}
\renewcommand{\P}{\mathbb{P}}
\newcommand{\Q}{\mathbb{Q}}
\newcommand{\dif}{\mathrm{d}}
\newcommand{\wti}{\widetilde}
\newcommand{\pb}{\mathbb{P}}
\newcommand{\vep}{\varepsilon}
\newcommand{\iden}{\mathbf{1}}
\newcommand{\half}{\frac{1}{2}}
\newcommand{\thalf}{\tfrac{1}{2}}
\newcommand{\R}{\mathbb{R}}
\newcommand{\sgn}{\operatorname{sgn}}
\newcommand{\Lc}{\mathcal{L}}
\newcommand{\ds}{\,\mathrm{d}s}
\newcommand{\dt}{\,\mathrm{d}t}
\newcommand{\F}{\mathcal{F}}
\newcommand{\Fc}{\mathcal{F}}
\newcommand{\eps}{\varepsilon}
\newcommand{\supp}{\operatorname{supp}}

\makeatother

\begin{document}
\begin{frontmatter}

\title{Root's barrier: Construction, optimality and applications to variance options}
\runtitle{Root's barrier: Construction, optimality and applications}

\begin{aug}
\author[A]{\fnms{Alexander M.~G.} \snm{Cox}\corref{}\ead[label=e1]{a.m.g.cox@bath.ac.uk}}
\and
\author[A]{\fnms{Jiajie} \snm{Wang}\ead[label=e2]{jiajie.wang@bath.edu}}
\runauthor{A.~M.~G. Cox and J. Wang}
\affiliation{University of Bath}
\address[A]{Department of Mathematical Sciences\\
University of Bath, Claverton Down\\
Bath BA2 7AY\\
United Kingdom\\
\printead{e1}\\
\phantom{E-mail:\ }\printead*{e2}}

\end{aug}

\received{\smonth{6} \syear{2011}}
\revised{\smonth{1} \syear{2012}}

%
\begin{abstract}
Recent work of Dupire and
Carr and Lee
has highlighted
the importance of understanding the Skorokhod embedding originally
proposed by Root for the model-independent hedging of
variance options. Root's work shows that there exists a \textit
{barrier} from which one may define a stopping time which solves
the Skorokhod embedding problem. This construction has the
remarkable property, proved by Rost, that it minimizes
the variance of the stopping time among all solutions.

In this work, we prove a characterization of Root's barrier in terms
of the solution to a variational inequality, and we give an
alternative proof of the optimality property which has an important
consequence for the construction of subhedging strategies in the
financial context.
\end{abstract}

%


\end{frontmatter}
%
\section{Introduction} \label{secIntroduction}

In this paper, we analyze the solution to the Skorokhod embedding
problem originally given by Root~\cite{Root69}, and generalized by
Rost~\cite{Rost76}. Our motivation for this is recent work connecting the
solution to this problem to questions arising in mathematical finance---specifically model-indepen\-dent
bounds for variance options---which has been observed by Dupire~\cite{dupire}, Carr and Lee \cite
{CarrLee10} and
Hobson~\cite{HobsonSurvey}. The financial motivation can be described as
follows: consider a (discounted) asset which has dynamics under the
risk-neutral measure
\[
\frac{\mathrm{d}S_t}{S_t} = \sigma_t \,\dif W_t,
\]
where the process $\sigma_t$ is not necessarily known. We are
interested in variance options, which are contracts where the payoff
depends on the realized quadratic variation of the log-price process:
specifically, we have
\[
\mathrm{d}(\ln S_t) = \sigma_t \,\dif W_t - \thalf\sigma_t^2 \,\dif t
\]
and therefore
\[
\langle \ln S \rangle_T = \int_0^T \sigma_t^2 \,\dif t.
\]
An option on variance is then an option with payoff $F(\langle \ln S
\rangle_T)$. Important examples include variance swaps, which pay the
holder $\langle \ln S \rangle_T -K$, and variance calls which pay the
holder $(\langle\ln S\rangle_T - K)_+$. We shall be particularly
interested in the case of a variance call, but our results will extend
to a wider class of payoffs. Let $\mathrm{d}X_t = X_t \,\dif\wti
{W}_t$ for
a suitable Brownian motion $\wti{W}_t$ and we can find a (continuous)
time change $\tau_t$ such that $S_t = \wti{X}_{\tau_t}$, and so
\[
\mathrm{d}\tau_t = \frac{\sigma_t^2 S_t^2}{S_t^2} \,\dif t.
\]
Hence
\[
(\wti{X}_{\tau_T}, \tau_T) = \biggl(S_T , \int_0^T
\sigma_u^2 \,\dif u\biggr) = (S_T ,\langle \ln S\rangle_T ).
\]
Now suppose that we know the prices of call options on $S_T$ with
maturity $T$, and at all strikes (recall that $\sigma_t$ is not
assumed known). Then we can derive the law of~$S_T$ under the
risk-neutral measure from the Breeden--Litzenberger formula. Call this
law $\mu$. This suggests that the problem of finding a lower bound on
the price of a variance call (for an unknown $\sigma_t$) is equivalent
to
%
%
\begin{equation} \label{eqrootbasic} \mbox{find a stopping time
$\tau$ to minimize $\E(\tau-K)_+$, subject to
$\Lc(\wti{X}_{\tau}) = \mu$.}\hspace*{-35pt}
\end{equation}
This is essentially the problem for which Rost has shown that the
solution is given by Root's barrier. [In fact, the result trivially
extends to payoffs of the form $F(\langle \ln S \rangle_T)$ where
$F(\cdot)$ is a convex, increasing function.]

In this work, our aim is twofold: first, to provide a proof that
Root's barrier can be found as the solution to a particular
variational inequality, which can be thought of as the generalization
of an obstacle problem; second, we show that the lower bound which
is implied by Rost's result can be enforced through a suitable hedging
strategy, which will give an arbitrage whenever the price of a
variance call trades below the given lower bound. To accomplish this
second part of the paper, we will give a novel proof of the optimality
of Root's construction, and from this construction we will be able to
derive a suitable hedging strategy.

The use of Skorokhod embedding techniques to solve model-independent
(or robust) hedging problems in finance can be traced back to
Hobson~\cite{Hobson98}. More recent results in this direction include
Cox, Hobson and Ob{\l}{\'o}j~\cite{CoxHobsonObloj08}, Cox and Ob{\l
}{\'o}j~\cite{CoxObloj11b} and
Cox and Ob{\l}{\'o}j~\cite{CoxObloj11}. For a comprehensive survey
of the literature on
the Skorokhod embedding problem, we refer the reader to
Ob{\l}{\'o}j~\cite{Obloj04}. In addition, Hobson \cite
{HobsonSurvey} surveys the
literature on the Skorokhod embedding problem with a specific emphasis
on the applications in mathematical finance.\vadjust{\goodbreak}

Variance options have been a topic of much interest in recent years,
both from the industrial point of view, where innovations such as the
VIX index have contributed to a large growth in products which are
directly dependent on quantities derived from the quadratic variation,
and also on the academic side, with a number of interesting
contributions in the literature. The academic results go back to work
of Dupire~\cite{Dupire93} and Neuberger~\cite{Neuberger94}, who
noted that a variance
swap---that is, a contract which pays $\langle\ln S\rangle_T$, can be
replicated model-independently using a contract paying the logarithm
of the asset at maturity through the identity (from It\^o's lemma)
%
%
\begin{equation} \label{eqloghedge}
\ln(S_T) - \ln(S_0) = \int_0^T \frac{1}{S_t} \,\dif S_t - \frac{1}{2}
\langle\ln S\rangle_T.
\end{equation}
More recently, work on options and swaps on volatility and variance,
(in a model-based setting) includes
Howison, Rafailidis and Rasmussen \cite
{HowisonRafailidisRasmussen04}, Broadie and Jain~\cite
{BroadieJain08} and
Kallsen, Muhle-Karbe and Voss~\cite{KallsenMuhle-KarbeVoss11}.
Other work
\cite{Keller-Ressel11,Keller-ResselMuhle-Karbe10} has considered
the differences between the theoretical payoff ($\langle\ln
S\rangle_T$) and the discrete approximation which is usually
specified in the contract
[$\sum_k\ln(S_{(k+1)\delta}/S_{k\delta})^2$]. Finally, several papers
have considered variants on the model-independent problems~\cite{CarrLee10,CarrLeeWu11,DavisOblojRaval10} or problems where
the modeling assumptions are fairly weak. This latter framework is of
particular interest for options on variance, since the markets for
such products are still fairly young, and so making strong modelling
assumptions might not be as strongly justified as it could be in a
well-established market.

The rest of this paper is structured as follows: in
Section~\ref{secRootsSolution} we review some known results and
properties concerning Root's barrier. In Section~\ref{secobs}, we
establish a connection between Root's solution and an obstacle
problem, and then in Section~\ref{secvarineq} we show that by
considering an obstacle problem in a more general analytic sense (as a
variational inequality), we are able to prove the equivalence between
Root's problem and the solution to a variational inequality. In
Section~\ref{secOptimality}, we give a new proof of the optimality
of Root's solution and in Section~\ref{secvaropt} we show how this
proof allows us to construct model-independent subhedges to give
bounds on the price of variance options.

\section{Features of Root's solution} \label{secRootsSolution}

Our interest is in Root's solution to the Skorokhod embedding
problem. Simply stated, for a process $(X_t)_{t \ge0}$, the Skorokhod
embedding problem is to find a stopping time $\tau$ such that $X_\tau
\sim\mu$. In this paper, we will consider first the case where $X_0
= 0$, and $X_t$ is a continuous martingale and a time-homogeneous
diffusion, and later the case where $X_0 \sim\nu$, is a centred,
square integrable measure. In such circumstances, it is natural to
restrict to the set of stopping times for which $(X_{t \wedge
\tau})_{t \ge0}$ is a uniformly integrable (UI) process. We will
occasionally call stopping times for which this is true \textit{UI
stopping times}.\vadjust{\goodbreak} In the case where $\mu$ is centered and has a
second moment and the underlying process $X$ is a Brownian motion (or
more generally, a diffusion and martingale with diffusion coefficient
$\sigma$ such that $\sigma^2\geq\vep$ for some strictly positive
constant $\vep$), this is equivalent to the fact that $\E\tau<
\infty$. For the case of a general starting measure, there is a
natural restriction on the measures involved, which is that we
require
%
%
\begin{equation}\label{eqnpotcond}
\infty>\mathrm{U}_{\nu}(x) := - \int_{\R} |y-x| \nu(\mathrm{d}y)
\ge- \int_{\R
} |y-x|
\mu(\mathrm{d}y) =: \mathrm{U}_{\mu}(x),
\end{equation}
for all $x \in\R$. This assumption implies that $m:=\int x\nu(\dif
x)=\int x\mu(\dif x)$; see Chacon~\cite{MR0501374}. By Jensen's inequality,
such a constraint is clearly necessary for the existence of a suitable
pair $\nu$ and $\mu$; further, by Rost~\cite{Rost71}, it is the only
additional constraint on the measures we will need to impose. We
shall write
%
%
\begin{equation}
\label{eqSmudefn}
S(\mu) = \{\tau\dvtx  \tau\mbox{ is a stopping time}, X_\tau\sim\mu, (X_{t
\wedge\tau})_{t \ge0} \mbox{ is UI}\}.
\end{equation}
There are a number of important papers concerning the construction of
Root's barrier. The first work to consider the problem is
Root~\cite{Root69}, and this paper proved the existence of a certain
Skorokhod embedding when $X_t$ is a Brownian motion. Specifically,
Root showed that if $X_t$ is a Brownian motion with $X_0 = 0$, and
$\mu$ is the law of a centered random variable with finite variance,
then there exists a stopping time $\tau$, which is the first hitting
time of a \textit{barrier}, which is defined as follows:
%
%
\begin{defn}[(Root's barrier)] \label{defrootbarrier} A closed subset
$B$ of $[-\infty,+\infty]\times[0,+\infty]$ is a \textit{barrier} if:
\begin{longlist}[(1)]
\item[(1)] $(x,+\infty)\in B$ for all $x\in[-\infty,+\infty]$;
\item[(2)] $(\pm\infty,t)\in B$ for all $t\in[0,\infty]$;
\item[(3)] if $(x,t)\in B$, then $(x,s)\in B$ whenever $s>t$.
\end{longlist}
\end{defn}

\begin{figure}

\includegraphics{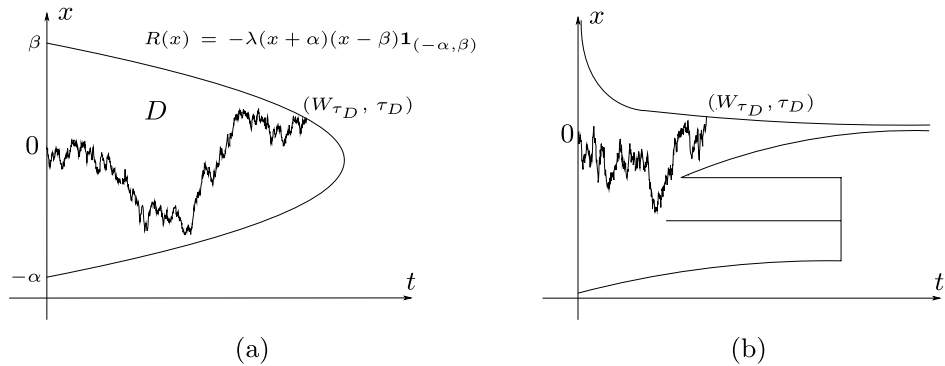}

\caption{Examples of Root's barriers: the
representation \textup{(a)} is an example of a ``nice''
barrier, where some explicit calculations can be made (see
Example~\protect\ref{exoptimal}); in \textup{(b)} we observe some
of the nastier features which a barrier may possess, including
spikes, corresponding to atoms of the distribution $\mu$ and
regions in which the barrier can be unbounded.}\label{figgraphRoot}
\end{figure}

We provide representative examples of barrier functions in
Figure~\ref{figgraphRoot}.
%
%

In a subsequent paper Loynes~\cite{Loynes70} proved a number of results
relating to barriers. From our perspective, the most important are,
first, that the barrier $B$ can be written as $B = \{(x,t) \dvtx  t \ge
R(x)\}$, where $R\dvtx \R\to[0,\infty]$ is a lower semi-continuous
function (with the obvious extensions to the definition to cover $R(x)
= \infty$); we will make frequent use of this representation. In
addition, Loynes~\cite{Loynes70}, Theorem~1, says that Root's solution is
essentially unique: if there are two barriers which embed the same
distribution with a UI stopping time, then their corresponding
stopping times are equal with probability one. The case where two
different barriers can occur are then only the cases where, say
$R(x_0) = 0$ for $x_0 > 0$, and then $R(x)$ is undetermined for all $x
> x_0$.

The other important reference for our purposes is
Rost~\cite{Rost76}. This work vastly extends the generality of the
results of Root and Loynes, and uses mostly potential-theoretic
techniques. Rost works in the generality of a Markov process $X_t$ on
a compact metric space $E$,\vadjust{\goodbreak} which satisfies the strong Markov property
and is right-continuous. Then Rost recalls (from an original
definition of Dinges~\cite{Dinges74} in the discrete setting) the
notion of
\textit{minimal residual expectation}:
%
%
\begin{defn}
A stopping time $\tau^* \in S(\mu)$ is of \textit{minimal residual
expectation} if, for each $t \in\R_+$, it minimizes the quantity
\[
\E(\tau- t)_+ = \E\int_{\tau\wedge t}^\tau\ds= \int_t^\infty
\P(\tau> s) \ds,
\]
over all $\tau\in S(\mu)$.
\end{defn}

Then Rost proves that [under \eqref{eqnpotcond}] there exists a
stopping time of minimal residual expectation
\cite{Rost76}, Theorem~1, and that the hitting time of any barrier
is of minimal residual expectation
\cite{Rost76}, Theorem~2. Finally, Rost also shows that the barrier
stopping times are, to a degree, unique~\cite{Rost76}, Corollary to Theorem~2. The relevant result for our purposes (where there
is a stronger form of uniqueness) is the corollary to Theorem~3
therein, which says that if $X_t$ is a process for which the one-point
sets are regular, then any stopping time of minimal residual
expectation is Root's stopping time. The class of processes for which
the one-point sets are regular include the class of time-homogenous
diffusions we consider.

Note that a stopping time is of minimal residual expectation if and
only if, for every convex, increasing function $F(t)$ (where, without
loss of generality, we take $F(0) = F_+'(0) = 0$), it minimizes the
quantity
\[
\E F(\tau) = \E\int_0^\infty(\tau-t)_+ F''(\dt),
\]
this fact being a consequence of the above representation.\vadjust{\goodbreak}

There are a number of important properties that the Root barrier
possesses. First, we note that, as a consequence of the fact that
$B$ is closed and the third property of
Definition~\ref{defrootbarrier}, the barrier is regular (i.e., if
we start at a point in the barrier, we will almost surely return to
the barrier instantly) for the class of processes we will consider
(time-homogeneous diffusions) this will have important analytical
benefits. Second, for a point $(x,t) \notin B$, we know that if the
stopped process at time~$t$ is at $x$, then we have not yet reached
the stopping time for the embedding. This will help in our
characterization of the law of the stopped process (Lemma~\ref{lemroot}).

In the rest of this paper, we will then say that a barrier is either a
lower semi-continuous function $R\dvtx \R\rightarrow\overline{\R}_{+}$,
with $R(0) \neq0$, or the complement of the corresponding connected
open set $D=\{(x,t)\dvtx  0<t<R(x)\} = \R\times(0,\infty) \setminus
B$. As noted above, by Loynes~\cite{Loynes70} this is equivalent to the
barrier as defined in Definition~\ref{defrootbarrier}. We will define
the hitting time of the barrier as: $\tau_{D}=\inf\{t > 0\dvtx
(X_{t},t)\notin D\}$. Note that the barrier $B$ is closed and regular,
so that $(X_{\tau_D},\tau_D) \in B$ and $\P^{(x,t)}(\tau_D = 0) = 1$
whenever $(x,t) \in B$, where $\P^{(x,t)}$ is the law of our diffusion
started at $x$ at time $t$.

Finally, we give some examples where the barrier function can be
explicitly calculated. We note that explicit examples appear to be the
exception, and in general are hard to compute. First, if $\mu$ is a Normal
distribution, we easily see that $R(x)$ is a constant. Second, if
$\mu$ consists of two atoms (weighted appropriately) at $a < 0 < b$
say, the corresponding barrier is
\[
R(x) =
\cases{
0, & \quad$x \notin(a,b),$ \vspace*{2pt}\cr
\infty,&\quad $x \in(a,b).$}
\]
In this example, observe that the function $R(x)$ is not unique:
we can choose any behavior outside $[a,b]$, and achieve the same
stopping time. Second, we note the that there are even more general
solutions to the Skorokhod embedding problem (without the uniform
integrability condition) since there are also barriers of the form
\[
R(x) =
\cases{
t_a, &\quad $x =a,$ \vspace*{2pt}\cr
t_b, &\quad $x = b,$ \vspace*{2pt}\cr
\infty,&\quad  $x \notin\{a,b\},$}
\]
%
which will embed the same law (provided $t_a, t_b > 0$ are chosen
suitably), but which do not satisfy the uniform integrability
condition. In
general, a barrier can exhibit some fairly nasty features: consider,
for example, the canonical measure on a middle third Cantor set $C$
(scaled so that it is on $[-1,1]$). Root's result tells us that there
exists a barrier which embeds this distribution, and clearly the
resulting barrier function must be finite only on the Cantor set;
however, the target distribution has no atoms, so that the ``spikes'' in
the barrier function can not be isolated (i.e., we must have
$\liminf_{y \uparrow x} R(y) = \liminf_{y \downarrow x} R(y) = R(x)$
for all $x \in(-1,1) \cap C$).

\section{Connecting Root's problem and an obstacle
problem} \label{secobs}

We now consider alternative methods for describing Root's barrier. We
will, in general, be interested in this question when our underlying
process $X_t$ is a solution to
%
%
\begin{equation} \label{eqX} \,\dif X_{t} = \sigma(X_{t})\,\dif W_{t},\qquad
X_0 \sim\nu,
\end{equation}
for a Brownian motion $(W_t)_{t \ge0}$, and we will introduce our
concepts in this general context. Initially, we assume that $\sigma\dvtx
\mathbb{R}\rightarrow\mathbb{R}$ satisfies, for some positive constant~$K$,
%
%
\begin{eqnarray}
\label{eqsde}
&| \sigma(x)-\sigma(y) | \leq K |x-y|;&
\\
\label{eqsde2}
&0 < \sigma^{2}(x) < K(1+x^2);&
\\
\label{eqhypo}& \sigma\mbox{ is smooth.}&
\end{eqnarray}
Recall that for the financial application we are interested in, we
want the specific case $\sigma(x) = x$ to be included. Clearly, this
case is currently excluded; however, we will show in
Section~\ref{secGBM} that the results can be extended to include this
case.

From standard results on SDEs, \eqref{eqsde} and \eqref{eqsde2}
imply that the unique strong solution $X^{a}$ of \eqref{eqX} with
$\nu= \delta_a$ is a strong Markov process with generator
$\frac{1}{2}\sigma^{2}\partial_{xx}$ for any initial value
$a\in\mathbb{R}$. Moreover, \eqref{eqhypo} implies that the operator
$L:=\frac{1}{2}\sigma^{2}\partial_{xx}-\partial_{t}$ is hypoelliptic;
see Stroock~\cite{Stroock08}, Theorem~3.4.1.

We will write Root's Skorokhod embedding problem as:
\begin{longlist}
\item[SEP$(\sigma,\nu,\mu)$:] Find a lower-semicontinuous function
$R(x)$ such that the domain $D = \{(x,t)\dvtx  0 < t < R(x)\}$ has
$X_{\tau_D} \sim\mu$, and $(X_{t \wedge\tau_D})_{t \ge0}$ is a UI
process, where $\nu$ is the initial law of $X_t$, and $\sigma$ the
diffusion coefficient.
\end{longlist}

Our aim is to show that the problem of finding $R$ is essentially
equivalent to solving an obstacle problem. Assuming that the relevant
derivatives exist, we shall show that the problem can be stated in the
following way:

\begin{longlist}
\item[OBS$(\sigma,\nu,\mu)$:] Find a function $u(x,t) \in
C^{1,1}(\R
\times\R_+)$ such that
\begin{subequations} \label{eqobs}
\begin{eqnarray}
&\mathrm{U}_{\nu}(x) = u(x,0),& \label{eqobsinit}\\
&0 \ge\mathrm{U}_{\mu}(x) - u(x,t) ,&\label{eqobsobs}\\
&\displaystyle 0 \ge\frac{\partial u}{\partial t}(x,t) - \half
\sigma(x)^2\frac{\partial^2u}{\partial x^2}(x,t),& \label
{eqobsdiff}\\
&\displaystyle\biggl(\frac{\partial u}{\partial t}(x,t) - \half
\sigma(x)^2\frac{\partial^2u}{\partial x^2}(x,t)\biggr)
\bigl(\mathrm{U}_{\mu}(x) -
u(x,t)\bigr) = 0,& \label{eqobscomp}
\end{eqnarray}
\end{subequations}
where \eqref{eqobsdiff} is interpreted in a distributional sense---that is, we require
\[
\int_\R\biggl(\phi(x) \frac{\partial u}{\partial t}(x,t) + \half
\sigma(x)^2\frac{\partial u}{\partial x}(x,t) \phi'(x)\biggr) \,\dif
x \le0
\]
whenever $\phi\in C_{K}^\infty$ is a nonnegative function. Condition
\eqref{eqobscomp} can be interpreted more generally as
requiring
\[
\frac{\partial u}{\partial t}(x,t) = \half\sigma(x)^2\frac{\partial
^2u}{\partial x^2}(x,t)
\]
in a distributional sense whenever $(\mathrm{U}_{\mu}(x) -
u(x,t)) \neq0$. However, this is an open set, and from the
hypoellipticity of the operator $L$, if this holds in a weak sense,
it will hold in a strong sense. Hence $\frac{\partial^2u}{\partial
x^2}(x,t)$ would be
continuous even if we were only to require \eqref{eqobscomp} to
hold in a distributional sense.\looseness=-1
\end{longlist}

In general, we do not expect $u$ to be sufficiently nice that we can
easily interpret all these statements, and one of the goals of this
paper is to give a generalization of {OBS}$(\sigma,\nu,\mu)$ that
will make sense more widely. Cases in which $u$ may not be expected to
be $C^{1,1}$ include the case where $\mu$ contains atoms (and
therefore $\mathrm{U}_{\mu}$ is not continuously differentiable). In
addition, we specify this problem in $C^{1,1}$ since, in general, we
would certainly not expect the second derivative to be continuous on
the boundary between the two types of behavior in \eqref{eqobscomp}.

\begin{thmm} \label{thmconstruction} Suppose $D$ is a solution to
{SEP}$(\sigma, \nu, \mu)$ and is such that
\[
u(x,t) = -\E|X_{t \wedge\tau_D} - x| \in C^{1,1}(\R
\times\R_+).
\]
Then $u$ solves {OBS}$(\sigma, \nu, \mu)$.
\end{thmm}

This gives an initial connection between {OBS}$(\sigma,\nu
,\mu)$
and {SEP}$(\sigma,\nu,\mu)$. We roughly expect solutions to Root's
problem to be the unique solutions to the obstacle problem (of course,
we do not currently know that such solutions exist or, when they do,
are unique). This suggests that we can attempt to solve the obstacle
problem to find the solution $D$ to Root's problem. In particular,
given a solution to {OBS}$(\sigma,\nu,\mu)$, we can now identify
$D$ as $D = \{(x,t)\dvtx  \mathrm{U}_{\mu}(x) < u(x,t), t >
0\}$.

%
\begin{lem}
\label{lemroot} For any $(x,t)\in D$,
$\pb(X_{t\wedge\tau_D}\in\dif x) = \pb(X_{t}\in
\,\dif
x,t<\tau_D)$.
\end{lem}

\begin{pf}
By the lower semi-continuity of $R$, since $(x,t)\in D$, there
exists \mbox{$h>0$} such that
\[
(x-h,x+h)\times[0,t+h)\subset D,
\]
and hence, for any $y\in(x-h,x+h)$, $R(y)>t$. On the other hand, if
$\tau_{D}\leq t$, we have
\[
R(X_{\tau_{D}}) \leq\tau_{D} \leq t,
\]
and hence, $X_{\tau_{D}}\notin(x-h,x+h)$. Therefore,
\begin{eqnarray*}
\pb(X_{t\wedge\tau_D}\in\dif x) & =&
\pb(X_{t}\in\dif x,
t<\tau_D)+\pb(X_{\tau_D}\in\dif x, t\geq\tau_D
) \\
& = &\pb(X_{t}\in\dif x,t< \tau_D).
\end{eqnarray*}
\upqed\end{pf}

%
\begin{lem}
\label{lemsmoothness}
The measure corresponding to $\Lc(X_{t}; t < \tau_D)$ has density
$p^D(x,t)$ with respect to Lebesgue on $D$, and the density is
smooth and satisfies
\[
\frac{\partial}{\partial t}p^D(x,t) = \half\frac{\partial
^2}{\partial x^2}[ \sigma(x)^2
p^D(x,t)].
\]
\end{lem}

This result appears to be standard, but we are unable to find concise
references. We give a short proof based on
\cite{RogersWilliamsVol2}, Section~V.38.5.

\begin{pf*}{Proof of Lemma \protect\ref{lemsmoothness}}
First note that, as a measure, $\Lc(X_{t}; t < \tau_D)$ is dominated
by the usual transition measure, so the density $p^D(x,t)$
exists.

Let $(x_0,t_0)$ be a point in $D$, and we can therefore find an
$\eps> 0$ such that $A = (x_0-\eps,x_0+\eps) \times
(t_0-\eps,t_0+\eps)$ satisfies $\bar{A} \subseteq D$. Then let $f$
be a smooth function, supported on $A$, and by It\^o's lemma,
\begin{eqnarray*}
f(X_{t \wedge\tau_D},t) &= & f(X_0,0) + \int_0^t \frac{\partial
f}{\partial x} (X_{s
\wedge\tau_D}, s) \,\dif X_s \\
& &{}+ \int_0^t \biggl( \half\sigma(X_{s \wedge
\tau_D})^2\frac{\partial^2}{\partial x^2} + \frac{\partial
}{\partial t} \biggr) f(X_{s \wedge
\tau_D}, s) \,\dif s.
\end{eqnarray*}
Since $f$ is compactly supported, taking $t > t_0 + \eps$, the two
terms on the left disappear, and the first integral term is a
martingale. Hence, taking expectations, and interchanging the order
of differentiation, we get
\[
\int_0^t \int p^D(y,s) \biggl( \half\sigma(y)^2 \frac{\partial
^2}{\partial x^2}
+ \frac{\partial}{\partial t} \biggr) f(y, s) \,\dif y \,\dif s = 0.
\]
Interpreting $p^D(y,s)$ as a distribution, we have
\[
\half\frac{\partial^{2}}{\partial x^{2}} [ \sigma(x)^2
p^D(x,t)] - \frac{\partial}{\partial t}
p^D(x,t) = 0,
\]
for $(x,t) \in A$, and since the heat operator is hypoelliptic, we
conclude that $p^D(x,t)$ is smooth in $A$ (e.g., Stroock~\cite{Stroock08}, Theorem~3.4.1).
\end{pf*}

We are now able to prove that any solution to Root's embedding problem
is a solution to the obstacle problem.

\begin{pf*}{Proof of Theorem~\protect\ref{thmconstruction}}
We first observe that $u(x,0) = -\E|X_0 - x|$, and $X_0 \sim\nu$,
so that $u(x,0) = -\int|y-x| \nu(\mathrm{d}y)$ and \eqref{eqobsinit}
holds. Second, since $(X_{t \wedge\tau_D})_{t \ge0}$ is a UI
process, by (conditional) Jensen's inequality,
\[
u(x,t) = -\E|x - X_{t \wedge\tau_D}| \ge- \E\bigl[ \E[
|x -
X_{\tau_D}| | \Fc_{t \wedge\tau_D} ]\bigr] =
\mathrm{U}_{\mu}(x),
\]
and \eqref{eqobsobs} holds.

We now consider \eqref{eqobsdiff}. Suppose $(x,t) \in D$, and note that
%
%
\begin{equation}\label{eqpotgrad}
\frac{\partial u}{\partial x} = 1- 2\P(X_{t \wedge\tau_D} < x),
\end{equation}
and therefore (in $D$) by Lemma~\ref{lemsmoothness} the function
$u$ has a smooth second derivative in $x$. Further, we
get
%
%
\begin{eqnarray}\label{eqDPDE}
\half\int_{0}^{t} \sigma(x)^2 \frac{\partial^{2}u}{\partial
x^{2}}(x,s) \,\dif s & = & -\int_0^t
\sigma(x)^2 p_D(x,s) \,\dif s \nonumber\\
& = & \lim_{\vep\downarrow0}\E\biggl[ \frac{
1}{ 2\vep}\int
_{0}^{t\wedge
\tau_{D}}\sigma(x)^2\iden_{[x-\vep<X_{s}<x+\vep]} \,\dif
s\biggr]
\nonumber
\\[-8pt]
\\[-8pt]
\nonumber
& = & -\E L_{t \wedge\tau_D}^x \\
& = & -\E|x-X_{t \wedge\tau_D}| + |x|,\nonumber
\end{eqnarray}
where $L^{x}_t$ is the local time of the diffusion at $x$. It
follows that $u$ satisfies \eqref{eqobsdiff} on $D$, and in fact
attains equality there. On the other hand, if $(x,t) \notin D$, it
follows from the definition of the barrier that if $\tau_{D}> t$,
the diffusion cannot cross the line $\{(x,s)\dvtx  s\geq t\}$ in the time
interval $[t,\tau_{D})$, and hence
\[
L_{t\wedge\tau_D}^x = L_t^x\iden_{\tau_D>t}+L_{\tau_D}^x
\iden_{\tau_D\leq t} = L_{\tau_D}^x\iden_{\tau_D>t}+L_{\tau_D}^x
\iden_{\tau_D\leq t} = L_{\tau_D}^x.
\]
Therefore, for $t\geq R(x)$,
\[
\E| x-X_{t\wedge\tau_{D}}|
= |x|+\E L^x_{t\wedge\tau_D}
= |x|+\E L^x_{\tau_D}
= \E|x-X_{\tau_{D}}|,
\]
%
%
where the last equality holds because $\tau_D$ is a UI stopping time.
So \eqref{eqobsobs} holds with equality when $(x,t) \notin D$. In
particular, we can deduce that either (if $(x,t) \in D$) we have
equality in \eqref{eqobsdiff}, or we have equality in
\eqref{eqobsobs}, in which case \eqref{eqobscomp} must hold. It
remains to show that \eqref{eqobsdiff} holds when $(x,t) \notin
D$. However, to see this, consider $(x,t) \notin D$, and note first
that $u(x,s) = u(x,t) = \mathrm{U}_{\mu}(x)$ whenever $s > t$, since
$(x,s) \notin D$. Hence $\frac{\partial u}{\partial t}(x,t) = 0$. It is
straightforward to check that $u(x,t)$ is concave in $x$, and therefore
that $\frac{\partial^{2}u}{\partial x^{2}}(x,t) \le0$, and \eqref
{eqobsdiff} also holds.
\end{pf*}

This result connects Root's problem and the obstacle problem under a
smoothness assumption on the function $u$. However, ideally we want a
one-to-one correspondence. We know from the results of Rost~\cite{Rost76}
that there always exists a solution to {SEP}$(\sigma,\nu,\mu)$,
and from Loynes~\cite{Loynes70} that the solution is unique. Our aim
is to
show that a similar combination of existence and uniqueness hold for
the corresponding analytic formulation. As already noted, we cannot
make a strong smoothness assumption on the function $u(x,t)$ as
required by {OBS}$(\sigma,\nu,\mu)$, and so we need a weaker
formulation of this problem. Generalizations of the obstacle problem
are well understood, and commonly called variational inequalities. In
the next section, we will reformulate the obstacle problem as a\vadjust{\goodbreak}
variational inequality, and we are able to state a problem for which
existence and uniqueness are known due to existing results.

\section{Root's barrier and variational inequalities} \label{secvarineq}

We now study the relation between Root's Skorokhod embedding problem
and a variational inequality. Our notation and definitions, and some
of the key results which we will use, come from
Bensoussan and Lions~\cite{BensoussanLions82}.

\subsection{Variational inequalities}
We begin with some necessary notation and results concerning
evolutionary variational inequalities. Given a constant $\lambda>0$
and a finite time $T>0$, we define the Banach spaces $H^{m,\lambda}
\subseteq L^2(\R)$ and $L^{2}(0,T;H^{m,\lambda})$ with
the norms
\begin{eqnarray*}
\|g\|^{2}_{H^{m,\lambda}} & =&
\sum_{k=0}^{m}\int_{\R}e^{-2\lambda|x|}\biggl| \frac{\partial^{k}
g}{\partial x^{k}}(x)
\biggr|^2\,\dif x; \\
\|w\|^{2}_{L^{2}(0,T;H^{m,\lambda})} & =&
\int_{0}^{T}\|w(\cdot,t)\|^{2}_{H^{m,\lambda}}\,\dif t,
\end{eqnarray*}
where the derivatives $\frac{\partial^{k} g}{\partial x^{k}}(x)$ are
to be interpreted as weak
derivatives---that is, $\frac{\partial^{k} g}{\partial x^{k}}(x)$
is defined by the
requirement that
\[
\int_{\R} \phi(x) \frac{\partial^{k} g}{\partial x^{k}}(x) \,\dif x
= (-1)^k \int_{\R} g(x)
\frac{\partial^{k} \phi}{\partial x^{k}}(x) \,\dif x,
\]
for all $\phi\in C_K^\infty(\R)$, and $C_{K}^{\infty}$ is the set of
compactly supported, smooth functions on $\R$. In particular, the
spaces $H^{m,\lambda}$ and $L^{2}(0,T;H^{m,\lambda})$ are Hilbert
spaces with respect to the obvious inner products. In addition,
elements of the set $H^{1,\lambda}$ can always be taken to be
continuous, and $C_{K}^{\infty}$ is dense in $H^{m,\lambda}$; see, for
example, Friedman~\cite{FriedmanGenFunc}, Theorem~5.5.20.

For functions $a(x,t), b(x,t) \in L^{\infty}(\R\times(0,T))$,
we define an operator
\[
a_{\lambda}(t;v,w) = \int_{\R} e^{-2\lambda|x|}
\biggl[a(x,t)\frac{\partial v}{\partial x}\frac{\partial w}{\partial
x} + b(x,t)\frac{\partial v}{\partial x} w
\biggr]\,\dif x,
\]
for $v,w\in L^2(0,T;H^{1,\lambda})$. Moreover if $\partial a/\partial
x$ exits, we define, for $v\in H^{2,\lambda}$,
\[
A(t)v = -\frac{\partial}{\partial x}\biggl(a(x,t)\frac{\partial
v}{\partial x}\biggr) +
\bigl(b(x,t)+2\lambda a(x,t) \sgn(x)\bigr)\frac{\partial
v}{\partial x}.
\]
And finally, for $v,w\in H^{0,\lambda}$,
\[
(v,w)_{\lambda} = \int_{\R}e^{-2\lambda|x|}vw\,\dif x,
\]
so that, for suitably differentiable test functions $\phi(x)$ and
$v\in H^{2,\lambda}$,
\[
(\phi,A(t)v)_\lambda= a_{\lambda}(t;v,\phi).
\]
Then we have the following restatement of Bensoussan and Lions~\cite{BensoussanLions82},
Theorem~2.2, and Section~2.15, Chapter~3:
\begin{thmm}
\label{thmB&L}
For any given $\lambda>0$ and $T>0$, suppose:
\begin{longlist}[(1)]
\item[(1)] $a,b,\frac{\partial a}{\partial t}$ are bounded on $\R
\times(0,T)$ with
$a(x,t) \geq\alpha$ a.e. in $\R\times(0,T)$ for some $\alpha>0$;
\item[(2)] $\psi, \frac{\partial\psi}{\partial t} \in
L^{2}(0,T;H^{1,\lambda}),\bar{v}\in H^{1,\lambda}, \bar{v} \ge\psi(0)$;
\item[(3)] the set
\begin{eqnarray*}
\mathcal{X} &:=& \biggl\{ w\in L^{2}(0,T;H^{1,\lambda})\dvtx
\frac{\partial w}{\partial t}\in
L^{2}(0,T;(H^{1,\lambda})^*),\\
&&\hspace*{91pt}w(t) \ge\psi(t) \mbox{ a.e. t in } [0,T] \biggr\}
\end{eqnarray*}
is nonempty, where $(H^{1,\lambda})^*$ denotes the dual space of
$H^{1,\lambda}$.
\end{longlist}

Then there exists a unique function $v$ such that:
\begin{eqnarray}
\label{eqsv1}
&\displaystyle v\in L^\infty(0,T;H^{1,\lambda}), \qquad \frac{\partial v}{\partial t}\in
L^2(0,T;H^{0,\lambda
});&\\
\label{eqsv2}
&\displaystyle\biggl(\frac{\partial v}{\partial t},w-v \biggr)_{\lambda}
+a_{\lambda}(t; v,
w-v) \geq0,&
\nonumber
\\[-8pt]
\\[-8pt]
\nonumber
&\hspace*{86pt}\forall w \in H^{1,\lambda}\mbox{ such that } w \ge
\psi(t) \mbox{ a.e. } t \in(0,T);&
\\
\label{eqsv3}
& v(\cdot,t) \ge\psi(t)\qquad \mbox{a.e. } t \in(0,T);&\\
\label{eqsv4}
& v(\cdot,0) = \bar{v}.&
\end{eqnarray}
Moreover, if $v\in L^{2}(0,T;H^{2,\lambda})$, then $v$ is a solution
to the obstacle problem: find $v\in L^{2}(0,T;H^{2,\lambda})$ such
that $v$ satisfies \eqref{eqsv3}, \eqref{eqsv4} and
\begin{eqnarray}
\label{eqvi1}
 \frac{\partial v}{\partial t} + A(t)v &\geq&0;\\
\label{eqvi3}
\biggl(\frac{\partial v}{\partial t} + A(t)v \biggr)(v-\psi) &=& 0,
\end{eqnarray}
almost everywhere in $\R\times(0,T)$.
\end{thmm}

\begin{pf}
For the most part, the theorem is a restatement of
Bensoussan and Lions~\cite{BensoussanLions82}, Theorem~2.2, and Section~2.15,
Chapter~3, where we have mapped $t \mapsto
T-t$, and $v \mapsto-v$.

We therefore only need to explain the last part of the result. If we
suppose $v\in L^2(0,T;H^{2,\lambda})$ and $\phi\in H^{1,\lambda}$,
we have
\begin{eqnarray*}
a_{\lambda}(t;v,\phi) & =& \int_{\R}e^{-2\lambda|x|} a(x,t)
\frac{\partial v}{\partial x} \,\dif\phi+ \int_{\R}e^{-2\lambda|x|}
\phi\biggl[
b(x,t) \frac{\partial v}{\partial x} \biggr]\,\dif x\\
&=&\biggl[e^{-2\lambda|x|} a(x,t) \frac{\partial v}{\partial x}\phi
\biggr]_{-\infty}^{\infty} +\int_{\R}e^{-2\lambda|x|} \phi\cdot
A(t)v \,\dif x,
\end{eqnarray*}
where the first term on the right-hand side vanishes since $v\in
L^{2}(0,t;H^{1,\lambda})$ and $\phi\in H^{1,\lambda}$. Therefore, by
\eqref{eqsv2}, for any $w\in H^{1,\lambda}$ such that $w \ge\psi$
a.e. in~$\R$,
\[
\biggl(\frac{\partial v}{\partial t}+A(t)v , w-v\biggr)_{\lambda}
\geq0\qquad
\mbox{a.e. }t.
\]
Taking, for example, $w = v + \phi$, for a positive test function
$\phi$, we conclude that~\eqref{eqvi1} holds. Moreover, let $w=\psi$
in the inequality above, we have
\[
\int_{\R}e^{-2\lambda|x|}\biggl(\frac{\partial v}{\partial t} +A(t)
v \biggr)(\psi-
v)\,\dif x \geq0.
\]
Then \eqref{eqvi3} follows from \eqref{eqsv3} and \eqref{eqvi1}.
\end{pf}

\subsection{Connection with Skorokhod's embedding problem}
\label{secsv&sep}
To connect our embedding problem {SEP}$(\sigma, \nu, \mu)$
with the
variational inequality, we need some assumptions on $\sigma$, $\mu$
and the starting distribution $\nu$. First, on
$\sigma\dvtx \R\rightarrow\R_+$, we still assume \eqref{eqsde} and
\eqref{eqhypo} hold. In addition, we assume that
%
%
\begin{equation}
\label{eqsde3}
\exists K >0\qquad \mbox{such that } \frac{1}{K} < \sigma< K \mbox{ on }
\R.
\end{equation}
On $\mu$ and $\nu$, we still assume that $\mathrm{U}_{\mu}(x)
\leq\mathrm{U}_{\nu}(x)$ to ensure the existence of a solution to
{SEP}$(\sigma,\nu,\mu)$.

Under these assumptions, we can specify the coefficients in the
evolutionary variational inequality, \eqref{eqsv4} and
\eqref{eqvi1}--\eqref{eqvi3}, to be
%
%
\begin{eqnarray}
\label{eqcoefficient}
a(x,t) &= &\frac{\sigma^2(x)}{2};\qquad b(x,t) =
\sigma(x)\sigma'(x)- \lambda\sigma^2(x) \sgn(x);
\nonumber
\\[-8pt]
\\[-8pt]
\nonumber
\psi(x,t) &=& \mathrm{U}_{\mu}(x);\qquad
\bar{v} = \mathrm{U}_{\nu}(x),
\end{eqnarray}
and then the corresponding operators are given by $A(t) =
-\frac{\sigma^2(x)}{2}\frac{\partial^2}{\partial x^2}$ and
\[
a_{\lambda}(t;v,w) = \int_{\R}e^{-2\lambda|x|}
\biggl[\frac{\sigma^2(x)}{2}\frac{\partial v}{\partial x}\frac
{\partial w}{\partial x}
+\bigl(\sigma(x)\sigma'(x) - \lambda\sigma^2(x) \sgn(x)
\bigr) \frac{\partial v}{\partial x} w\biggr]\,\dif x.
\]

We write the evolutionary variational inequality as:
\begin{longlist}
\item[VI$(\sigma,\nu,\mu)$:] Find a function $v\dvtx \R\times[0,T]
\rarr
\R$ satisfying \eqref{eqsv1}--\eqref{eqsv4}, where all the coefficients
are given in \eqref{eqcoefficient}.
\end{longlist}
We also have a stronger formulation, that is:
\begin{longlist}
\item[SVI$(\sigma,\nu,\mu)$:] For given $T>0$, we seek a function $v$,
in a suitable space, such that \eqref{eqsv3}--\eqref{eqvi3} hold, where
all the coefficients are given in \eqref{eqcoefficient}.
\end{longlist}

Our main result is then to show that finding the solution to {SEP}$(\sigma,\nu,\mu)$ is equivalent to finding a (and hence the
unique) solution to {VI}$(\sigma,\nu,\mu)$:
\begin{thmm}\label{thmsv&sep}
Suppose \eqref{eqsde}, \eqref{eqhypo} and \eqref{eqsde3} hold,
and let $T>0$.
Also, let~$D$ and $v$ be the solutions to {SEP}$(\sigma,\nu,\mu)$ and {VI}$(\sigma,\nu,\mu)$,
respectively. Define $u(x,t) :=-\E^\nu|x-X_{t\wedge\tau_D}|$ and
$D^{T}$ by
%
%
\begin{equation}\label{eqDT}
D^{T} := \{(x,t)\in\R\times[0,T]; v(x, t)> \psi(x,t)\}.
\end{equation}
Then we have $D^{T}=D\cap\R\times[0,T]$, and for all
$(x,t) \in\R\times[0,T]$,
\[
u(x,t) = v(x,t).
\]
Moreover, if $u \in L^2(0,T;H^{2,\lambda})$, then $u$ is also the
solution to {SVI}$(\sigma, \nu, \mu)$.
\end{thmm}

\begin{pf}
Let $\lambda>0$ be fixed, and suppose $D$ is a solution to SEP$(\sigma,\nu,\mu)$. We need to show $u$ is a solution to
VI$(\sigma,\nu,\mu)$. First note that $\mathrm{U}_{\mu}(x)+|x|$ is
continuous on $\R$, and converges to $0$ as $x\rightarrow\pm\infty$,
and hence is bounded. So $x\mapsto\mathrm{U}_{\mu}(x)+|x|\in
L^\infty(0,T;H^{0,\lambda})$, and then $\mathrm{U}_{\mu}(x) \in
L^\infty(0,T;H^{0,\lambda})$. Similarly, $\mathrm{U}_{\nu}(x) \in
L^\infty(0,T;H^{0,\lambda})$. Since $0 \ge\mathrm{U}_{\nu}(x) \ge u(x,t)
\ge
\mathrm{U}_{\mu}(x)$ for all $t \in[0,T]$, we have $u\in
L^\infty(0,T;H^{0,\lambda})$. By \eqref{eqpotgrad}, we also have
$|\frac{\partial u}{\partial x}|\leq1$ since $u$ is the
potential of some
probability distribution. Therefore we have $u\in
L^{\infty}(0,T;H^{1,\lambda})$. By Lemma~\ref{lemsmoothness} and the
fact that $u$ is constant (in time) outside $D$, $| \frac
{\partial u}{\partial t}
| \leq\sigma^2p^\nu(x,t)$ a.e. on $\R\times[0,T]$ where
$p^\nu(x,t)$ is the transition density of the diffusion process~$X$
starting from $\nu$. Then by standard Gaussian estimates (e.g.,
Stroock~\cite{Stroock08}, Theorem~3.3.11), we know there exists some
constant $A>0$, depending only on~$K$, such that
\begin{eqnarray*}
&&\biggl\|\frac{\partial u}{\partial t} \biggr \|_{L^2(0,T;H^{0,\lambda
})} \\
&&\qquad \leq\int_\R\int_{0}^{T}\int_{\R} \frac{A}{1\wedge t} \exp
\biggl\{-2 \biggl(At-\frac{(x-y)^2}{At} \biggr)^{-}
-2 \lambda|x| \biggr\} \,\dif x \,\dif t\, \nu(\mathrm{d}y)\\
&&\qquad= \int_\R\int_{0}^{T}\frac{A}{1\wedge t} \int_{y-At}^{y+At}
e^{-2\lambda|x|} \,\dif x \,\dif t\, \nu(\mathrm{d}y)\\
&&\qquad\quad {}+ \int_\R\int_{0}^{T} \frac{Ae^{2At}}{1 \wedge
t} \int_{\R\setminus(y-At,y+At)} \exp
\biggl\{-\frac{2(x-y)^2}{At}
-2\lambda|x| \biggr\}\,\dif x \,\dif t\, \nu(\mathrm{d}y)\\
&&\qquad\leq\int_\R\int_{0}^{T} \frac{A}{1\wedge t} \int_{-At}^{At}
e^{-2\lambda|x|} \,\dif x \,\dif t\, \nu(\mathrm{d}y)\\
&&\qquad\quad {}+\int_\R\int_{0}^{T} \frac{Ae^{2At}}{1\wedge t}
\int_{\R\setminus(y-At,y+At)} \exp\biggl\{-\frac{2(x-y)^2}{At}
\biggr\} \,\dif x \,\dif t\, \nu(\mathrm{d}y)\\
&&\qquad= \frac{A}{\lambda} \int_{0}^{T} \frac{1}{1\wedge t}
(1-e^{-2\lambda At}) \,\dif t + 2A \int_{0}^{T}
\frac{e^{2At}}{1\wedge t} \int_{At}^{\infty} \exp
\biggl\{-\frac{2z^{2}}{At}\biggr\} \,\dif z \,\dif t\\
&&\qquad\leq\frac{A}{\lambda} \int_{0}^{T} \frac{2 A\lambda t}{1
\wedge t} \,\dif t + \frac{A^{3/2} \pi^{1/2}}{\sqrt{2}}
\int_{0}^{T} \frac{e^{2At} \sqrt{t}}{1 \wedge t} \,\dif t <
\infty,
\end{eqnarray*}
where we have applied H\"older's inequality in the first line to
get
\[
\biggl| \frac{\partial u}{\partial t}\biggr|^2 = \biggl| \int_\R
p(t,y,x) \nu(\mathrm{d}
y)\biggr|^2 \le\int_\R p(t,y,x)^2 \nu(\mathrm{d}y).
\]
So $\frac{\partial u}{\partial t} \in L^{2}(0,T;H^{0,\lambda})$, and
we have shown
\eqref{eqsv1} holds.

By the same arguments used in the proof of
Theorem~\ref{thmconstruction}, \eqref{eqsv3} and \eqref{eqsv4}
hold. Now we consider \eqref{eqsv2}. We begin by observing that,
for any $\phi\in C^{\infty}_K$, if we write $\mu_t(\mathrm{d}x)$
for the
law of $X_{t \wedge\tau_D}$, we have
\begin{eqnarray}\label{equmeas}
\int_\R\frac{\partial\phi}{\partial x} \frac{\partial u}{\partial
x} \,\dif x & = &\int_\R\frac{\partial\phi}{\partial x}
\bigl( 1 - 2\P(X_{t \wedge\tau_D} \le x)\bigr) \,\dif x \nonumber\\
& =& -2 \int_\R\int_\R\frac{\partial\phi}{\partial x} \mathbf
{1}_{\{y \le x\}} \mu_t(\mathrm{d}y)
\,\dif x
\nonumber
\\[-8pt]
\\[-8pt]
\nonumber
& =& 2 \int_\R\phi(y) \mu_t(\mathrm{d}y) \\
& =& 2 \E[\phi(X_{t \wedge\tau_D})].\nonumber
\end{eqnarray}
In addition, for any $w\in H^{1,\lambda}$, we can find a sequence
$\{\phi_{n}\}\subset C^{\infty}_{K}$ such that
%
%
\begin{equation}\label{eqphiseq}
\lim_{n\rightarrow\infty}\bigl\|\phi_{n}-\bigl(w-u(\cdot,t)\bigr)\bigr\|
_{H^{1,\lambda}}
= 0.
\end{equation}
Moreover, $e^{-\lambda|x|}u(x,t)$ is bounded, and if
$e^{-\lambda|x|} w$ is also bounded, then we can, in addition, find
a sequence $\{\phi_{n}\}\subset C^{\infty}_{K}$ such that
$e^{-2\lambda|x|} \phi_n(x) \ge-K'$ for some constant $K'$
independent of $n$. For any $n$, we therefore have
%
%
\begin{eqnarray}
\label{eqip}
\qquad\int_{\R}e^{-2\lambda|x|} \frac{\sigma^2}{2} \frac{\partial
u}{\partial x}
\frac{\partial\phi_n}{\partial x} \,\dif x& =& -\int_{\R}
e^{-2\lambda|x|}
\bigl(\sigma\sigma' -\lambda\sigma^2 \sgn(x) \bigr)
\frac{\partial u}{\partial x} \phi_n \,\dif x
\nonumber
\\[-8pt]
\\[-8pt]
\nonumber
&& {} + \int_{\R} e^{-2\lambda|x|} \phi_n \sigma^2
\mu_{t}(\mathrm{d}x).
\end{eqnarray}

On the other hand, since $\partial u/\partial t$ vanishes outside
$D$, and, using the same arguments as \eqref{eqDPDE} (which still
hold on account of Lemma~\ref{lemsmoothness}), is equal to $-
\sigma(x)^2 p^D(x,t)$, we have, for almost every $t\in[0,T]$
%
%
\begin{eqnarray}
\label{eqvi}
&&\int_{\R} e^{-2\lambda|x|} \phi_n \frac{\partial u}{\partial t}
\,\dif x +
\int_{\R}e^{-2\lambda|x|} \phi_n \sigma^2 \mu_{t}(\mathrm{d}x)
\nonumber
\\[-8pt]
\\[-8pt]
\nonumber
&&\qquad= \int_{\R\setminus D_{t}} e^{-2\lambda|x|} \phi_n \sigma^2 \mu
_{t}(\mathrm{d}x),
\end{eqnarray}
where $D_{s}:=\{x \in\R\dvtx  (x,s)\in D\}$. By \eqref{eqip} and
\eqref{eqvi},
\begin{eqnarray*}
&&\biggl(\frac{\partial u}{\partial t},\phi_n\biggr)_{\lambda
}+a_{\lambda}(t;u,\phi
_n)\\
&&\qquad=  \int_{\R}e^{-2\lambda|x|} \biggl[ \frac{\partial u}{\partial
t}\phi_{n}
+\frac{\sigma^2}{2} \frac{\partial u}{\partial x}\frac{\partial
\phi_{n}}{\partial x} +
\bigl(\sigma\sigma'-\lambda\sigma^2 \sgn(x) \bigr)
\frac{\partial u}{\partial x} \phi_{n} \biggr] \,\dif x\\
&&\qquad=  \int_{\R} e^{-2\lambda|x|} \phi_n \frac{\partial u}{\partial
t} \,\dif x
+ \int_{\R} e^{-2\lambda|x|} \phi_n \sigma^2 \mu_{t}(\mathrm
{d}x) \\
&&\qquad=  \int_{\R\setminus D_{t}} e^{-2\lambda|x|} \phi_n \sigma^2
\mu_{t}(\mathrm{d}x),
\end{eqnarray*}
for almost every $t\in[0,T]$. Now suppose initially we have
$e^{-\lambda|x|} w$ bounded, and choose a sequence $\phi_n$ as
above. Then we can let $n \to\infty$ and apply Fatou's lemma and
the fact that $u=\psi$ on $\R\setminus D_{t}$ and $w \ge\psi$ to get
\[
-\biggl(\frac{\partial u}{\partial t},w-u\biggr)_{\lambda
}+a_{\lambda}(t;u,w-u)
= \int_{\R\setminus D_{t}} e^{-2\lambda|x|} (w-\psi) \sigma^2
\mu_{t}(\mathrm{d}x) \geq0,
\]
for almost every $t \in[0,T]$. So \eqref{eqsv2} holds when
$e^{-\lambda|x|} w$ is bounded. The general case follows from
noting that $\max\{w,-N\}$ converges to $w$ in $H^{1,\lambda}$. We can
conclude that $u$ is a solution to VI$(\sigma,\nu,\mu)$. In addition, the final statement of the
theorem now follows from Theorem~\ref{thmB&L}.

Conversely, suppose that we have already found the solution to
VI$(\sigma,\nu,\mu)$, denoted by $v(x,t)$. By
Theorem~\ref{thmB&L} and the preceding argument, we have
\[
-\E^\nu|x-X_{t\wedge\tau_{D}}| = v(x,t),
\]
when $(x,t)\in\R\times[0,T]$. Finally, we need only note (from
\eqref{eqDPDE}, and the line above) that whenever $(x,t) \in D$, we
have $u(x,t) > \psi(x,t)$, and hence $D^{T}=D\cap\R\times[0,T]$.
\end{pf}

\begin{rmk}\label{rmkdifflambda}
The constant $\lambda$ which appears in
the variational inequality can now be seen to be unimportant: if we
consider two positive numbers $\lambda<\lambda^{\ast}$, then by
Theorem~\ref{thmB&L}, there exist $v$ and $v^{\ast}$ satisfying
\eqref{eqsv1}--\eqref{eqsv4} with the parameters $\lambda$ and
$\lambda^{\ast}$, respectively. According to
Theorem~\ref{thmsv&sep},
\[
u(x,t) = v(x,t) = v^{\ast}(x,t),
\]
so $v=v^{\ast}$. Therefore, the description of Root's barrier
by the strong variational inequality is not affected by the choice
of the parameter $\lambda>0$. We do, however, need $\lambda> 0$,
since this assumption is used, in, for example, \eqref{eqip}, to
ensure we can
integrate by parts.
\end{rmk}

\begin{rmk} \label{rmkoptstop}
As noted in Bensoussan and Lions~\cite{BensoussanLions82}, and which
is well known, one
can connect the solution to the variational inequality
VI$(\sigma,\nu,\mu)$ to the solution of a particular optimal
stopping problem. In our context, the function~$v$ which arises in\vadjust{\goodbreak}
the solution to {VI}$(\sigma,\nu,\mu)$ is also the function
which arises from solving the problem
%
%
\begin{equation} \label{eqoptstop}
v(x,t) = \sup_{\tau\le t}\E^{x} \bigl[ \mathrm{U}_{\mu}(X_\tau)
\mathbf{1}_{\{\tau< t\}} + \mathrm{U}_{\nu}(X_\tau) \mathbf{1}_{\{
\tau= t\}}\bigr].
\end{equation}
This seems a rather interesting observation, and at one level
extends a number of connections known to exist between solutions to
the Skorokhod embedding problem, and solutions to optimal stopping
problems (e.g., Peskir~\cite{Peskir98}, Ob{\l}{\'o}j \cite
{Obloj07} and
Cox, Hobson and Ob{\l}{\'o}j~\cite{CoxHobsonObloj08}).

What is rather interesting, and appears to differ from these other
situations, is that the above examples are all cases where the same
stopping time is both a Skorokhod embedding, and a solution to the
relevant optimal stopping problem. In the context here, we see that
the optimal stopping problem is \textit{not} solved by Root's stopping
time. Rather, the problem given in \eqref{eqoptstop} runs
``backwards'' in time: if we keep $t$ fixed, then the solution to
\eqref{eqoptstop} is
\[
\tau_D = \inf\{s \ge0\dvtx  (X_s,t-s) \notin
D\}\wedge t.
\]
In addition, our connection between these two problems is only
through the analytic statement of the problem: it would be
interesting to have a probabilistic explanation for the
correspondence.
\end{rmk}

\begin{rmk} \label{rknoncentred} The above ideas also allow us to
construct alternative embeddings which fail to be uniformly
integrable. Consider using the variational inequality to construct
the domain $D$ in the manner described above, but with the function~$\psi$
chosen to be $\mathrm{U}_{\mu}(x) - \alpha$, for some
$\alpha>0$.
By \eqref{eqoptstop}, one can check that the solution to the
variational inequality is a decreasing function with respect to $t$,
and hence, $B= D^\complement$ is a barrier, which is nonempty, so
that $\tau_D < \infty$ a.s., and the functions $u(x,t)$ and
$v(x,t)$ defined in Theorem~\ref{thmsv&sep} agree (e.g., by
taking bounded approximations to $D$). In particular, $\lim_{t \to
\infty} u(x,t) = \mathrm{U}_{\mu}(x)-\alpha$. Since $X_{t \wedge
\tau_D}$ is no longer uniformly integrable, we cannot simply infer
that this holds in the limit, but we can consider for example
\[
u(x,t) - u(z,t) = - \E[ |X_{t \wedge\tau_D}-x| - |X_{t
\wedge\tau_D}-z| ]
\]
which is a bounded function. Taking the limit as $t \to\infty$, we
can deduce that
\[
- \E[ |X_{\tau_D}-x| - |X_{\tau_D}-z| ] =
\mathrm{U}_{\mu}(x) - \mathrm{U}_{\mu}(z).
\]
From this expression, we can divide through by $(x-z)$ and take the
limit as $x \downarrow z$ to get $2\P(X_{\tau_D} >z) - 1$. The law
of $X_{\tau_D}$ now follows.

Note also that there is no reason that the distribution above needed
to have the same mean as $\nu$, and this can lead to constructions
where the means differ. In general, these constructions will not
give rise to a uniformly integrable embedding, but if we take two
general (integrable) distributions, there is a natural choice, which
is to find the smallest $\alpha\in\R$ such that $\mathrm{U}_{\nu}(x)
\ge\mathrm{U}_{\mu}(x) - \alpha$. In such a case, we conjecture
that the
resulting construction would be \textit{minimal} in the sense that
there is no other construction of a stopping time which embeds the
same distribution, and is almost surely smaller. See
Monroe~\cite{Monroe72} and Cox~\cite{Cox05} for further details regarding
minimality.
\end{rmk}

\subsection{Geometric Brownian motion} \label{secGBM}

An important motivating example for our study is the financial
application of Root's solution described in the \hyperref[secIntroduction]{Introduction}. In both
\cite{dupire} and~\cite{CarrLee10}, the case $\sigma(x)=x$ plays a
key role in both the pricing and the construction of a hedging
portfolio. However, in the previous section, we only discussed the
relation between Root's construction and variational inequalities
under the assumptions \eqref{eqsde}, \eqref{eqhypo} and
\eqref{eqsde3}, where the last assumption is not satisfied by~$\sigma$ in this special case.

In this section, we study this special case: $\sigma(x) = x$, so that
$X_t$ is a geometric Brownian motion. In addition, we will assume that
the process is strictly positive, so that $\nu$ and $\mu$ are
supported on
$(0,\infty)$. We therefore consider the Skorokhod embedding problem
{SEP}$(\sigma,\nu,\mu)$ with starting distribution $\nu$, where
$\nu$
and $\mu$ are integrable probability distributions satisfying
%
%
\begin{eqnarray}
\label{eqmu3}
\supp(\mu) &\subset&(0,\infty), \qquad\supp(\nu) \subset(0,\infty),
\nonumber
\\[-8pt]
\\[-8pt]
\nonumber
\mathrm{U}_{\mu}(x)&\leq&\mathrm{U}_{\nu}(x) \quad\mbox{and} \quad\int
x^2\,\dif\nu
<\infty.
\end{eqnarray}
We recall from \eqref{eqnpotcond} that this implies, in particular,
that the means of $\mu$ and $\nu$ agree.

The solution to the stochastic differential equation
\[
\dif X_{t} = X_{t}\,\dif W_{t},\qquad X_0 = x_0
\]
is the geometric Brownian motion
$x_0 \exp\{W_t-t/2\}$, and, for $y>0$, the transition density of the
process is
%
%
\begin{equation}
\label{eqexpdensity}
p_t(y,x) := \frac{1}{x}\frac{1}{\sqrt{2\pi t}}\mathbf{1}_{\{x>0\}}
\exp\biggl\{-\frac{(\ln x-\ln y+t/2)^2}{2t}\biggr\}.
\end{equation}

By analogy with Theorem~\ref{thmconstruction}, if $D$ is the solution
to {SEP}$(\sigma,\nu,\mu)$, then we would expect
\[
\frac{\partial u}{\partial t} = \frac{x^2}{2}\frac{\partial
^{2}u}{\partial x^{2}} \qquad\mbox{on } D;\qquad u(x,t)
= \mathrm{U}_{\mu}(x)\qquad \mbox{on } \R\times(0,\infty)\setminus D;
\]
where $u$ is defined as before by $u(x,t) =
-\E|x-X_{t\wedge\tau_D}|$. However, if we follow the arguments in
Section~\ref{secsv&sep}, we find that we need to set $a(x,t)=x^2/2$
in {VI}$(\sigma,\nu,\mu)$, which would not satisfy the first
condition of Theorem~\ref{thmB&L}. To avoid this we will perform a
simple transformation of the problem. We set
\[
v(x,t) = u(e^{ x},t), \qquad (x,t)\in\R\times[0,T].
\]
Define the operator $A(t):=-\frac{1}{2}\frac{\partial^2}{\partial x^2}
+\frac{1}{2}\frac{\partial}{\partial x}$; then we have, when
$(e^x,t)\in D$,
%
%
\begin{equation}
\label{eqspecialvi}
\frac{\partial v}{\partial t}+A(t)v = 0.
\end{equation}

We state our main result of this section as follows:
\begin{thmm} \label{thmmainGBM}
Suppose $\sigma(x)=x$ on $(0,\infty)$ and $\mu$ and $\nu$ satisfy
\eqref{eqmu3}. Moreover, assume $D$ solves SEP$(\sigma,\nu,\mu)$, and $u(x,t)
:=-\E|x-X_{t\wedge\tau_D}|$. Then $v(x,t) :=u(e^x,t)$ is the unique
solution to \eqref{eqsv1}--\eqref{eqsv4} where we set
%
%
\begin{eqnarray}\label{eqscoefficient}
 a(x,t)& =& \thalf;\qquad b(x,t) = \thalf-\lambda\cdot\sgn(x);\qquad
\psi(x,t) = \mathrm{U}_{\mu}(e^x);
\nonumber
\\[-8pt]
\\[-8pt]
\nonumber
 \bar{v} &=&
\mathrm{U}_{\nu}(e^{x});\qquad \lambda> \thalf.
\end{eqnarray}
\end{thmm}

\begin{pf}
Much of the proof will follow the proof of Theorem~\ref{thmsv&sep}.
As before, \eqref{eqsv3} and \eqref{eqsv4} are clear. In
addition, we note that $\psi-e^x$ is continuous and converges to $0$
as $x\rightarrow\infty$ and converges to $\mathrm{U}_{\mu}(0) <
\infty$
as $x \to-\infty$, so $x\mapsto\psi-e^x\in
L^\infty(0,T;H^{0,\lambda})$. Hence $\psi\in
L^{\infty}(0,T;H^{0,\lambda})$ since we have $\lambda>\half$. Thus,
$v\in L^{\infty}(0,T;H^{0,\lambda})$. Moreover, we can easily see
$|\partial v/\partial x|$ is bounded by $e^x$. Therefore, $v\in
L^{\infty}(0,T;H^{1,\lambda})$ when $\lambda> \half$. On the other
hand, since $|\partial v/\partial t|$ is bounded by $e^{2x}\int
p_{t}(y,e^x) \nu(\mathrm{d}y)$, we have, by H\"older's inequality,
\[
\biggl|\frac{\partial v}{\partial t}\biggr|^2 \leq\int_{\R
_+}\frac{1}{2\pi t}
\exp\biggl\{-\frac{(x-\ln y+t/2)^2}{t}+2x\biggr\}\nu(\dif y),
\]
and hence,
\begin{eqnarray*}
&&\biggl\|\frac{\partial v}{\partial t}\biggr\|_{L^2(0,T;H^{0,\lambda
})} \\
&&\qquad\leq
\int_{\R_+}\int_{0}^{T}\int_{\R}\frac{e^{-2\lambda
|x|}}{2\pi t} \exp\biggl\{-\frac{(x-\ln
y+t/2)^{2}}{t}+2x\biggr\}\,\dif x\,\dif t\, \nu(\mathrm{d}y)\\
&&\qquad\leq\int_{\R_+}\int_{0}^{T}\int_{\R}\frac{1}{2\pi
t}\exp\biggl\{-\frac{(x-\ln y-t/2)^{2}}{t} +2\ln y \biggr\}
\,\dif x\,\dif t\, \nu(\mathrm{d}y)\\
&&\qquad\leq\int_{\R_+}y^2 \nu(\mathrm{d}
y)\int_{0}^{T}\frac{1}{2\sqrt{\pi t}}\,\dif t < \infty.
\end{eqnarray*}
Therefore \eqref{eqsv1} is verified.

Using \eqref{equmeas}, for $\phi\in C_K^{\infty}$ we get
\begin{eqnarray}\label{eqspecialdistr}
\int_\R\biggl( \frac{\partial\phi}{\partial x}(x) + \phi(x)
\biggr) \frac{\partial v}{\partial x} \,\dif x
& =& \int_0^\infty\frac{\partial}{\partial y} [ \phi(\ln
(y))y]
\frac{\partial u}{\partial x}(y,t) \,\dif y
\nonumber
\\[-8pt]
\\[-8pt]
\nonumber
& =& 2 \E[\phi(\ln(X_{t \wedge\tau_D})) X_{t \wedge
\tau_D}],
\end{eqnarray}
and so we define the measure $\nu_t$ by
\[
\int\phi(x) \nu_t(\mathrm{d}x) = \E[\phi(\ln(X_{t \wedge
\tau_D})) X_{t \wedge\tau_D}].
\]
Now take any $w\in H^{1,\lambda}$, and take $\{\phi_n\}\subset
C^{\infty}_{K}$ satisfying \eqref{eqphiseq}. By \eqref
{eqspecialvi} and~\eqref{eqspecialdistr}, similar arguments to those used in the
proof of Theorem~\ref{thmsv&sep} give
\[
\int_{\R} e^{-2\lambda|x|} \frac{\partial v}{\partial x} \biggl(
\half
\frac{\partial\phi_n}{\partial x} + \half\phi_n - \lambda\cdot
\sgn(x)\biggr)
\,\dif x
= \int_{\R}e^{-2\lambda|x|} \phi_n \nu_{t}(\mathrm{d}x),
\]
and
\[
\int_{\R} e^{-2\lambda|x|} \frac{\partial v}{\partial t} \phi
_n\,\dif x
+ \int_{\R}e^{-2\lambda|x|}\phi_n \nu_{t}(\mathrm{d}x)
= \int_{\R\setminus\wti{D}_{t}}e^{-2\lambda|x|}\phi_n \nu
_{t}(\mathrm{d}x),
\]
for almost all $t \in[0,T]$, where $\wti{D}_{t}:=\{x\in\R\dvtx
(e^x,t)\in
D\}$. Thus, for almost every $t \in[0,T]$,
\begin{eqnarray*}
&&\biggl(\frac{\partial v}{\partial t},\phi_n\biggr)_\lambda+
a_{\lambda}(t;v,\phi_n) \\
&&\qquad = \int_\R\biggl( \frac{\partial v}{\partial t} \phi_n +
\half\frac{\partial\phi_n}{\partial x} \frac{\partial v}{\partial
x} + \biggl( \half- \lambda
\cdot\sgn(x)\biggr) \phi_n \frac{\partial v}{\partial x} \biggr)
\,\dif x \\
&&\qquad = \int_{\R\setminus\wti{D}_{t}}e^{-2\lambda|x|}\phi_n \nu
_{t}(\mathrm{d}x).
\end{eqnarray*}
Finally, following the same arguments as in the proof of
Theorem~\ref{thmsv&sep}, we conclude~\eqref{eqsv2}
holds. Therefore $v$ is a solution to \eqref{eqsv1}--\eqref{eqsv4}
with coefficients determined by \eqref{eqscoefficient}. The
uniqueness is clear since it is easy to check the coefficients
defined in \eqref{eqscoefficient} satisfy the conditions in
Theorem~\ref{thmB&L}.
\end{pf}

\section{Optimality of Root's solution} \label{secOptimality}

For a given distribution $\mu$, Rost~\cite{Rost76} proves that Root's
construction is optimal in the sense of ``minimal residual
expectation.'' It is easy to check that this is equivalent to the
slightly more general problem
\begin{eqnarray*}
&&\mbox{minimize }  \E[ F(\tau)] \\
&&\mbox{subject to:}\qquad \mathcal{L}(X_{\tau}) = \mu;\\
&&\phantom{\mbox{subject to:}\qquad }\tau\mbox{ is a UI stopping time.}
\end{eqnarray*}
Here we assume $\mu$ is a given integrable and centered distribution,
$X$ is the diffusion process defined by \eqref{eqX}, where the
diffusion coefficient $\sigma$ satisfies
\eqref{eqsde}--\eqref{eqhypo}, with initial distribution
$\mathcal{L}(X_{0}) = \nu$, and $F$ is a given convex, increasing
function with right derivative $f$ and $F(0) = 0$.

Our aim in this section is twofold. First, since Rost's original
proof relies heavily on notions from potential theory, to give a proof
of this result using probabilistic techniques. Second, we shall be
able to give a ``pathwise inequality'' which encodes the optimality in\vadjust{\goodbreak}
the sense that we can find a submartingale $G_t$, and a function~$H(x)$ such that
%
%
\begin{equation}
\label{eqpathwiseineq} F(t) \ge G_t + H(X_t)
\end{equation}
and such that, for $\tau_D$, equality holds in~\eqref{eqpathwiseineq} and $G_{t \wedge\tau_D}$ is a UI
martingale. It then follows that $\tau_D$ does indeed minimize $\E
F(\tau)$ among all solutions to the Skorokhod embedding problem. The
importance of \eqref{eqpathwiseineq} is that we can characterize the
submartingale $G_t$, which will correspond in the financial setting
to a dynamic trading strategy for constructing a sub-replicating
hedging strategy for call-type payoffs on variance options.

We first define the key functions $G(x,t)$ and $H(x)$, where the
submartingale in~\eqref{eqpathwiseineq} is $G_t = G(X_t,t)$, and give
key results concerning these functions.

We suppose that we have solved Root's problem for the given
distributions, and hence have our barrier $B = D^{\complement}$.
Define the
function
%
%
\begin{equation} \label{eqMdefn}
M(x,t) = \E^{(x,t)}f(\tau_{D}),
\end{equation}
where $\tau_D$ is the corresponding Root stopping time. In the
following, we shall assume
%
%
\begin{equation}
\label{eqMassumption}
M(x,t) \mbox{ is locally bounded on } \R\times\R_+.
\end{equation}
We suppose also (at least initially) that
\eqref{eqsde}--\eqref{eqhypo} and \eqref{eqsde3} still hold. Note
that $M(x,t)$ now has the following important properties. First, since
$f$ is right-continuous (it is the right derivative of $F$), $M(x,t) =
f(t)$ whenever $(x,t) \notin D$ and $t>0$. In addition, since $f$ is
increasing, for all $x$ and $t$ we have $M(x,t) \ge f(t)$.

Now define a function $Z(x)$ by
%
%
\begin{equation} \label{eqZdefn}
Z(x) = 2 \int_0^x \int_0^y \frac{M(z,0)}{\sigma^2(z)} \,\dif z
\,\dif y.
\end{equation}
So in particular, we have $Z''(x) = 2\frac{M(x,0)}{\sigma^2(x)}$, and
$Z(x)$ is a convex function. Define also
%
%
\begin{equation} \label{eqGdefn}
G(x,t) = \int_0^t M(x,s) \,\dif s - Z(x),
\end{equation}
and
%
%
\begin{equation}
\label{eqHdefn}
H(x) = \int_0^{R(x)} \bigl(f(s) - M(x,s)\bigr) \ds+ Z(x),
\end{equation}
where $R(x)$ is the barrier function. Two key results concerning these
functions are then:
%
%
\begin{prop} \label{propGFineq} We have, for all $(x,t) \in\R\times
\R_+$,
%
%
\begin{equation}
\label{eqGFineq}
G(x,t) + H(x) \le F(t).
\end{equation}
\end{prop}

And also:\vadjust{\goodbreak}
%
%
\begin{lem} \label{lemGmartsubmart} Suppose that $f$ is bounded, and
for any $T > 0$,
%
%
\begin{equation}
\label{eqqvbddcond}
\E\biggl[ \int_0^T Z'(X_s)^2 \sigma(X_s)^2 \,\dif s \biggr] <
\infty, \qquad \E Z(X_0) < \infty.
\end{equation}
Then the process
%
%
\begin{equation} \label{eqGmart} G(X_{t\wedge\tau_D},t \wedge
\tau_D)\qquad \mbox{is a martingale,}
\end{equation}
and
%
%
\begin{equation}\label{eqGsubmart}
G(X_{t},t) \qquad\mbox{is a submartingale.}
\end{equation}
\end{lem}

Using these results, we are able to prove the following theorem, which
gives us Rost's result regarding the optimality of Root's
construction.
\begin{thmm} \label{thmoptimality} Suppose $D$ solves {SEP}$(\sigma, \mu, \nu)$,
and equations \eqref{eqMassumption} and
\eqref{eqqvbddcond} hold. Then
%
%
\begin{equation} \label{eqtauineq}
\E F(\tau_D) \le\E F(\tau)
\end{equation}
whenever $\tau$ is a stopping time such that $X_{\tau} \sim\mu$.
\end{thmm}

\begin{pf}
We begin by considering the case where $\E\tau_D < \infty, \E\tau
< \infty$ and $f$ is bounded. Since $Z(x)$ is convex, by the
Meyer--It\^{o} formula (e.g.,
Protter~\cite{Protter05}, Theorem~IV.71),
\[
Z(X_t) = Z(X_0) + \int_0^t Z'(X_r) \,\dif X_r + \half\int_0^t
Z''(X_r) \sigma^2(X_r) \,\dif r.
\]
By \eqref{eqqvbddcond} and the fact that $f$ is bounded (and hence
also $M(X_s,0)$ is bounded), we get
\[
\E Z(X_{t \wedge\tau}) = \E Z(X_0) + \E\int_0^{t \wedge\tau}
M(X_s,0) \,\dif
s \le f(\infty) \E\tau+ \E Z(X_0).
\]
Applying Fatou's lemma, we deduce that for any stopping time $\tau$
with finite expectation, $Z(X_\tau)$ is
integrable. Moreover for such a stopping time, by convexity, $Z(X_{t
\wedge\tau}) \le\E[Z(X_{\tau}) | \Fc_t]$, and so,
by Lemma~\ref{lemGmartsubmart}, $G(X_{t \wedge
\tau},t\wedge\tau)$ is a submartingale which is bounded below by
a UI martingale, and bounded above by $f(\infty) \tau$. It follows
that $\E G(X_{t \wedge\tau},t\wedge\tau) \to\E G(X_{\tau},\tau)$
as $t \to\infty$. The same arguments hold when we replace $\tau$ by
$\tau_D$.

Since $R(X_{\tau_D}) \le\tau_D$ and if $t \in[R(x),\infty)$, then
$\tau_D = t, \P^{(x,t)}$-a.s., so that $M(X_{\tau_D},s) = f(s)$
for $s \ge\tau_D$, we have
\begin{eqnarray} \label{eqGFequal}
&& G(X_{\tau_D},\tau_D) +  \int_0^{R(X_{\tau_D})} \bigl(f(s) -
M(X_{\tau_D},s)\bigr) \ds+ Z(X_{\tau_D}) \nonumber\\
&&\qquad = \int_0^{\tau_D}
M(X_{\tau_D},s) \,\dif s + \int_0^{R(X_{\tau_D})} \bigl(f(s) -
M(X_{\tau_D},s)\bigr) \ds
\nonumber
\\[-8pt]
\\[-8pt]
\nonumber
&&\qquad = \int_0^{\tau_D}
M(X_{\tau_D},s) \,\dif s + \int_0^{\tau_D} \bigl(f(s) -
M(X_{\tau_D},s)\bigr) \ds\\
&&\qquad = \int_0^{\tau_D} f(s) \,\dif s = F(\tau_D).\nonumber
\end{eqnarray}
On the other hand, since $X_{\tau_D} \sim X_{\tau}$, and observing
that $G(X_{\tau_D},\tau_D)$ and $F(\tau_D)$ are integrable, so too
is $H(X_{\tau_D})$, and
\[
\E H(X_{\tau_D}) = \E H(X_{\tau}).
\]

In addition, by Lemma~\ref{lemGmartsubmart} and the limiting
behavior deduced above, we have
\[
\E G(X_{\tau_D},\tau_D) = \E G(X_{0},0) \le\lim_{t \to\infty} \E
G(X_{t \wedge
\tau},t \wedge\tau) = \E G(X_{\tau},\tau).
\]
Putting these together, we get
\begin{eqnarray*}
\E F(\tau_D) & =& \E[G(X_{\tau_D},\tau_D) + H(X_{\tau_D})
]
\\
& \le&\E
[G(X_{\tau},\tau) + H(X_\tau) ]\\
& \le&\E F(\tau).
\end{eqnarray*}

We now consider the case where at least one of $\tau$ or $\tau_D$
has infinite expectation. Note that if $F(\cdot) \not\equiv0$, then
there is some $\alpha, \beta\in\R$ with $\beta>0$ such that $F(t)
\ge\alpha+ \beta t$, and hence we cannot have $\E\tau= \infty$
or $\E\tau_D = \infty$ without the corresponding term in
\eqref{eqtauineq} also being infinite. The only case which need
concern us is the case where $\E\tau< \infty$, but $\E\tau_D =
\infty$. Note, however, that $\tau_D$ remains UI, so $\E[X_{t \wedge
\tau_D} | \Fc_t] = X_t$. In addition, from the arguments applied
above, we know $Z(X_\tau)$ is integrable, and since $X_\tau\sim
X_{\tau_D}$, so too is $Z(X_{\tau_D})$. Then $H(X_\tau)$ and
$H(X_{\tau_D})$ are both bounded above by an integrable random
variable, so their expectations are well defined (although possibly
not finite), and equal. Then, as above, $-\E[Z(X_{\tau_D})|\Fc_t]
\le-Z(X_{t \wedge\tau_D}) \le G(X_{t \wedge\tau_D}, t \wedge
\tau_D)$. We can deduce that $\E G(X_{\tau_D},\tau_D) \le\lim_{n
\to\infty} \E G(X_{t \wedge\tau_D},t \wedge\tau_D) = G(X_0,0)
\le\E G(X_\tau,\tau)$. The remaining steps follow as previously,
and it must follow that in fact $\E F(\tau_D) \le\E F(\tau)$, which
contradicts the assumption that $\E\tau< \infty$ and $\E\tau_D =
\infty$.

To observe that the result still holds when $f$ is unbounded,
observe that we can apply the above argument to $f(t) \wedge N$, and
$F_N(t) = \int_0^s f(s) \wedge N \,\dif s$ to get $\E F_N(\tau_D) \le
\E F_N(\tau)$, and the conclusion follows on letting $N\to\infty$.
\end{pf}

We now turn to the proofs of our key results:

\begin{pf*}{Proof of Proposition~\protect\ref{propGFineq}}
If $t \le R(x)$, then the left-hand side of \eqref{eqGFineq} is
\[
\int_0^t f(s) \,\dif s + \int_t^{R(x)} \bigl(f(s) - M(x,s)\bigr) \,\dif s = F(t) -
\int_t^{R(x)} \bigl(M(x,s)-f(s)\bigr) \,\dif s,
\]
and we know $M(x,s) \ge f(s) \ge0$, so that the inequality holds.

Now consider the case where $R(x) \le t$. Then the left-hand side of
\eqref{eqGFineq} becomes
\[
\int_{R(x)}^t M(x,s) \,\dif s + \int_0^{R(x)} f(s) \,\dif s =
\int_{R(x)}^t f(s) \,\dif s + \int_0^{R(x)} f(s) \,\dif s = F(t).
\]
\upqed\end{pf*}

\begin{pf*}{Proof of Lemma~\protect\ref{lemGmartsubmart}}
We begin by noting that $Z(x)$ is convex, and therefore the
Meyer--It\^{o} formula (e.g., Protter~\cite{Protter05}, Theorem~IV.71)
gives
\[
Z(X_t) - Z(X_s) = \int_s^t Z'(X_r) \,\dif X_r + \half\int_s^t
Z''(X_r) \sigma^2(X_r) \,\dif r.
\]
It follows from \eqref{eqqvbddcond} that the first integral is a
martingale. So we get
\[
\E[ Z(X_t) - Z(X_s)| \Fc_s] = \int_s^t
\E[M(X_r,0)|\Fc_s] \,\dif r,\qquad
s \le t.
\]

In addition, since $M(x,t) \ge f(t)$ and $f(t)$ is increasing, for
$r,u\geq0$ by the strong Markov property, writing $\wti{X}$ for an
independent stochastic process with the same law as $X$ and
$\wti{\tau}_D$ for the corresponding hitting time of the barrier, we
have
\begin{eqnarray*}
\E^{(x,r)}[f(\tau_{D})|\mathcal{F}_{r+u}] & =&
\iden_{\tau_{D}>r+u} \E^{(x,r)}
[f(\tau_{D})|\mathcal{F}_{r+u}] \\
&&{}+ \iden_{\tau_{D}\leq
r+u} \E^{(x,r)}[f(\tau_{D})|\mathcal{F}_{r+u}] \\
&\leq&\iden_{\tau_{D}>r+u}\E^{(X^{x}_{u},r+u)}[f(\wti{\tau}_D)]
+\iden_{\tau_D\leq r+u} f(r+u)\\
&\leq& M(X^{x}_{u},r+u).
\end{eqnarray*}
When $r=0$, we have $\E^{(x,0)}[f(\tau_D)|\mathcal{F}_u]\leq
M(X^{x}_{u},u)$. For $s,u \in[0,t]$,
\begin{eqnarray}\label{eqMsubmart1}
\E[M(X_{t}, u) | \Fc_s] &=& \E^{X_s}M(\wti{X}_{t-s},u) \nonumber\\
& \ge&\E^{(X_{s}, u-(t-s))}[f(\wti{\tau}_D)] \\
& \ge& M\bigl(X_{s}, u-(t-s)\bigr),\nonumber
\end{eqnarray}
when $u \ge t-s$. On the other hand, if $u < t-s$,
\begin{eqnarray} \label{eqMsubmart2}
\E[ M(X_{t}, u) | \Fc_s] & =&
\E\bigl[\E^{(X_{t-u},0)}[M(\wti{X}_u,u)]|\Fc_s\bigr] \nonumber\\
&\geq&\E\bigl[\E^{(X_{t-u},0)}[f(\wti{\tau}_D)]|\Fc_s\bigr]
\\
& \ge&\E[M(X_{t-u},0)|\Fc_s].\nonumber
\end{eqnarray}

Then we can write
\begin{eqnarray*}
\E[ G(X_t,t) | \Fc_s] & =& \int_0^t \E[ M(X_t,u) |
\Fc_s] \,\dif u - \E[ Z(X_t) | \Fc_s] \\[-2pt]
& =& G(X_s,s) + \int_0^t \E[ M(X_t,u) | \Fc_s] \,\dif u -
\int_0^s M(X_s,u) \,\dif u \\[-2pt]
& &{}- \E[ Z(X_t)-Z(X_s) | \Fc_s] \\[-2pt]
& \ge& G(X_s,s) + \int_0^{t-s} \E[ M(X_{t-u},0) |
\Fc_s] \,\dif u - \int_0^s M(X_s,u) \,\dif u \\[-2pt]
&&{} -
\int_s^t \E[ M(X_{u},0) |
\Fc_s] \,\dif u + \int_{t-s}^t M(X_s,s-t+u) \,\dif u \\[-2pt]
& \ge& G(X_s,s) + \int_s^{t} \E[ M(X_{u},0) | \Fc_s]
\,\dif u - \int_s^t \E[ M(X_{u},0) | \Fc_s] \,\dif u \\[-2pt]
&&{}+
\int_{0}^s M(X_s,u) \,\dif u -\int_0^s M(X_s,u) \,\dif u \\[-2pt]
& \ge& G(X_s,s).
\end{eqnarray*}
Where we have used \eqref{eqMsubmart1} and \eqref{eqMsubmart2} in
the third line.

On the other hand, on $\{\tau_D \ge s\}$, from the definition of
$M(x,t)$ and the Markov property, we get
%
%
\begin{equation} \label{eqMmart1} \E[ M(X_{t \wedge\tau_D},
t \wedge\tau_D - u) | \Fc_s] = M(X_s, s-u)
\end{equation}
when $u \le s$, and
%
%
\begin{equation} \label{eqMmart2} \E[ M(X_{t \wedge\tau_D},
t \wedge\tau_D - u) | \Fc_u] = M(X_u, 0)
\end{equation}
when $u \in[s,t \wedge\tau_D]$. Then a similar calculation to
above gives, for $s \le\tau_D$,
\begin{eqnarray*}
&&\E [ G(X_{t\wedge\tau_D},t\wedge\tau_D) | \Fc_s] \\[-2pt]
&&\qquad = \E\biggl[ \int_0^{t\wedge\tau_D} M(X_{t\wedge\tau_D},{t\wedge
\tau_D}-u) \,\dif u \Big|
\Fc_s\biggr] - \E[ Z(X_{t\wedge\tau_D}) | \Fc_s] \\[-2pt]
&&\qquad = \int_0^s M(X_s,s-u)\,\dif u + \E\biggl[ \int_s^{t\wedge\tau_D}
M(X_{t\wedge\tau_D},{t\wedge\tau_D} - u) \,\dif u \Big| \Fc_s\biggr]
\\[-2pt]
&&\qquad\quad{} - Z(X_s) - \E\biggl[ \int_s^{t\wedge\tau_D} M(X_u,0)
\,\dif u \Big| \Fc_s\biggr]
\\[-2pt]
&&\qquad = \E\biggl[ \int_s^{t} \E[ M(X_{t\wedge\tau_D},{t\wedge
\tau_D} - u) - M(X_u,0) | \Fc_u] \mathbf{1}_{\{u \le\tau_D\}}
\,\dif u \Big |\Fc_s\biggr]
\\[-2pt]
&&\qquad\quad + G(X_s,s) \\[-2pt]
&&\qquad = G(X_s,s),
\end{eqnarray*}
where we have used \eqref{eqMmart1} and \eqref{eqMmart2}.\vadjust{\goodbreak}
\end{pf*}

\begin{rmk}
Note that the fact that our choice of $D$ given in the solution is
the domain $D$ which arises in solving Root's embedding problem is
only used in Theorem~\ref{thmoptimality} to enforce the lower
bound. In fact, we could choose any barrier $B$, and $D=B^\complement$
as our
domain, and this would result in a lower bound, with corresponding
functions $G$ and $H$. The choice of Root's barrier gives the
optimal lower bound, in that we can attain equality for some
stopping time. In this context, it is worth recalling the lower
bounds given by Carr and Lee~\cite{CarrLee10}, Proposition~3.1---here a lower
bound is given which essentially corresponds to choosing the domain
with $R(x) = Q$, for a constant $Q$. The arguments given above show
that similar constructions are available for any choice of $R$, and
the optimal choice corresponds to Root's construction.
\end{rmk}

\begin{rmk} \label{rmkbndsigisid}
Although the preceding section is written for a diffusion on $\R$,
it is not hard to check that the case where $\sigma(x) = x$ can also
be included without many changes. In this setting, we need to
restrict the space variable to the space $(0,\infty)$ (so we assume
that $\tau_D < \infty$ a.s.), and consider a starting distribution
which is also supported on $(0,\infty)$, and with a corresponding
change to \eqref{eqMassumption}.
\end{rmk}

\begin{figure}
\centering
\begin{tabular}{@{}cc@{}}

\includegraphics{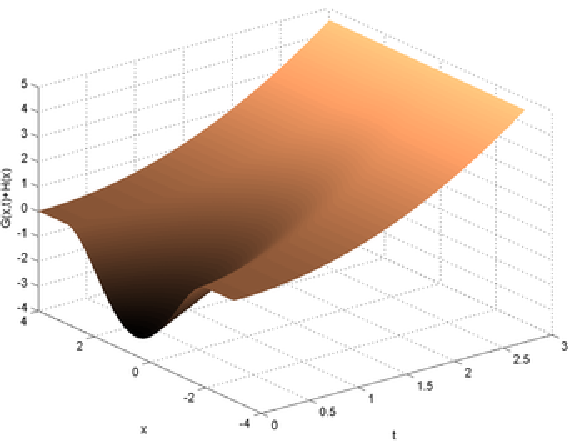}
 & \includegraphics{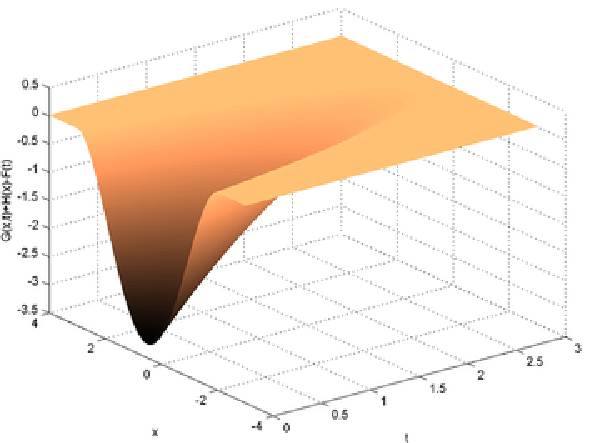}\\
\footnotesize{(a)} & \footnotesize{(b)}\\
\end{tabular}
\caption{We give graphical representations of some of
the relevant quantities derived in Example~\protect\ref{exoptimal}, for
$\alpha= 2, \beta= 3$ and $\lambda= 1/2$. In
\textup{(a)} we see $G(x,t) + H(x)$, which is a lower bound
for $F(t)$, and in \textup{(b)} we see the difference
$G(x,t) + H(x) - F(t)$, which is indeed negative.}\label{figOptEx}
\end{figure}

We end this section with a brief example which illustrates some of the
relevant quantities.
\begin{example} \label{exoptimal} Suppose we take Root's barrier
$D:=\{(x,t)\dvtx t<R(x)\}$ with the boundary function $R(x) = -\lambda
(x+\alpha)(x-\beta)\iden_{(-\alpha,\beta)}, $ where
$\lambda,\alpha,\beta>0$; see Figure~\ref{figgraphRoot}a. Given a
standard Brownian motion $W$ and Root's stopping time $ \tau_D = \inf
\{t>0\dvtx  t\geq R(W_t)\}, $ define $\mu:=\Lc(W_{\tau_D})$. Let
$F(t) = t^2/2$, and we will see $\E[F(\tau_D)]\le\E[F(\tau)]$ for
any UI
stopping time $\tau$ such that $W_\tau\sim\mu$.

For $(x,t)\in\R\times\R_+$, define $M(x,t)=\E^{(x,t)}[\tau_D]$.
Then if $t\geq R(x)$, $M(x,t)=t$. If $0\leq t<R(x)$, since
$\tau_D=\lambda(W_{\tau_{D}}+\alpha) (W_{\tau_{D}}-\beta)$, using
It\^o's formula, we can compute $M(x,t)$ to be
\[
M(x,t) = \frac{\lambda}{1+\lambda}
[ t-(x+\alpha)(x-\beta)]\qquad
\mbox{for }0\leq t <R(x).
\]
Defining $G,H,Z$ as in \eqref{eqZdefn}--\eqref{eqHdefn}, we get
the explicit expressions
\begin{eqnarray*}
Z(x)&=& \frac{\lambda}{6 (1+\lambda)} \cdot
\cases{
-\beta^4 - 2 \alpha\beta^3 + (2\beta^3 + 6 \alpha
\beta^2)x, & \quad $x \ge\beta,$\vspace*{2pt}\cr
-x^4 - 2(\alpha-\beta)x^3 + 6 \alpha\beta x^2, &\quad $x \in
(-\alpha,\beta),$\vspace*{2pt}\cr
-\alpha^4 - 2 \alpha^3 \beta- (2\alpha^3 +6 \alpha^2
\beta) x, & \quad $x \le-\alpha,$}
%
\\
G(x,t)&=&
\cases{
\displaystyle\frac{\lambda}{1+\lambda}\biggl[\frac{t^2}{2}
-t(x+\alpha)(x-\beta)\biggr]-Z(x), &\quad $\mbox{if }
0\leq t<R(x),$\vspace*{2pt}\cr
\displaystyle\frac{R^{2}(x)}{2(1+\lambda)} +\half t^{2}-Z(x),&\quad $\mbox{if } t\geq
R(x),$}
\\
H(x)&=&-\frac{R^2(x)}{2(1+\lambda)}+Z(x) .
\end{eqnarray*}
It is easy to check directly that $G(W_t,t)$ is a submartingale, and
that it is a martingale up to the stopping time $\tau_D$. We also
can check that \eqref{eqGFineq} holds here:
\[
G(x,t)+H(x)-F(t)=
\cases{
-\displaystyle\frac{[ R(x)-t ]^2}{2(1+\lambda)},
&\quad $\mbox{if } 0\leq t<R(x),$\vspace*{2pt}\cr
0,&\quad $\mbox{if } t\geq R(x).$}
\]
Therefore, for any UI stopping time $\tau$ such that
$\mathcal{L}(W_{\tau})=\mu=\mathcal{L}(W_{\tau_D})$,
%
%
\begin{eqnarray}
\label{eqoptimality}
\qquad \E[F(\tau)]
&\geq&\E[G(W_{\tau},\tau)+H(W_\tau)]\geq\E
[G(W_{\tau
_{D}},\tau_{D})]
+\E[H(W_{\tau_D})]
\nonumber
\\[-8pt]
\\[-8pt]
\nonumber
&=&\E[F(R(W_{\tau_{D}}))]
+\E\biggl[\int_{R(W_{\tau_{D}})}^{\tau_{D}} M(W_{\tau_{D}},s)\,\dif
s\biggr] =\E[F(\tau_{D})],
\end{eqnarray}
which shows the optimality of Root's stopping
time. Figure~\ref{figOptEx} illustrates some of the relevant
functions derived here.
\end{example}
%
%

\section{Financial applications} \label{secvaropt}

We now turn to our motivating financial problem: consider an asset
price $S_t$ defined on a complete probability space $(\Omega,
\Fc, ( \Fc_t)_{t \ge0},  \P)$, with
%
%
\begin{equation} \label{eqSdefn}
\frac{\mathrm{d}S_t}{S_t} = r_t \,\dif t + \sigma_t \,\dif W_t
\end{equation}
under some probability measure $\Q\sim\P$, where $\P$ is the
objective probability measure, and $W_t$ a $\Q$-Brownian motion. In
addition, we suppose $r_t$ is the risk-free\vadjust{\goodbreak} rate which we require to
be known, but which need not be constant. In particular, let~$r_t,
\sigma_t$ be locally bounded, predictable processes so that the
integral in \eqref{eqSdefn} is well defined, and so $S_t$ is an It\^o
process. We suppose that the process $\sigma_t$ is not known (or more
specifically, we aim to produce conclusions which hold for all
$\sigma_t$ in the class described). Specifically, we shall suppose:
\begin{assumption} \label{assPrice}
The asset price process, under some probability measure $\Q\sim\P$,
is the solution to the SDE \eqref{eqSdefn}, where $r_t$ and
$\sigma_t$ are locally bounded, predictable processes.
\end{assumption}

In addition, we need to make the following assumptions regarding the
set of call options, which are initially traded:
\begin{assumption} \label{assCalls} We suppose that call options with
maturity $T$, and at all strikes $\{K\dvtx  K \ge0\}$ are
traded at time $0$, and the prices, $C(K)$, are assumed to be
known. In addition, we suppose call-put parity holds, so that the
price of a put option with strike $K$ is $P(K) = e^{-\int_0^T r_s
\,\dif s}K -S_0 + C(K)$. We make the additional assumptions that
$C(K)$ is a continuous, decreasing and convex function, with $C(0)=
S_0$, $C_+'(0) = -e^{-\int_0^T r_s \,\dif s}$ and $C(K) \to0$ as $K
\to\infty$.
\end{assumption}

Many of these notions can be motivated by arbitrage concerns; see, for
example, Cox and Ob{\l}{\'o}j~\cite{CoxObloj11}. That there are
plausible situations in
which these assumptions do not hold can be seen by considering models
with bubbles (e.g.,~\cite{CoxHobson05}), in which call-put parity
fails, and $C(K) \not\to0$ as $K \to\infty$. Let us define $B_t =
e^{\int_0^t r_s \,\dif s}$, and make the assumptions above. Following
the perspective that the prices correspond to expectations under $\Q$,
the implied law of $B_T^{-1}S_T$ (which we
will denote $\mu$) can be recovered by the Breeden--Litzenberger
formula~\cite{BreedenLitzenberger78},
%
%
\begin{equation} \label{eqBL}
\mu((K,\infty)) = \Q^*\bigl(B_T^{-1}S_T \in(K,\infty)\bigr) = -2 B_T C_+'(B_TK).
\end{equation}
Here we have used $\Q^*$ to emphasize the fact that this is only an
\textit{implied} probability, and not necessarily the distribution under
the actual measure $\Q$. From \eqref{eqBL} we deduce that
$\mathrm{U}_{\mu}(x) = S_0 - 2C(B_Tx) - x$, giving an affine mapping
between the function $\mathrm{U}_{\mu}(x)$ and the call prices. We do not
impose the condition that the law of $B_T^{-1} S_T$ under $\Q$ is
$\mu$, we merely note that this is the law implied by the traded
options. We also do not assume anything about the price paths of the
call options: our only assumptions are their initial prices, and that
they return the usual payoff at maturity. It can now also be seen that
the assumption that $C_+'(0) = -B_T^{-1}$ is equivalent to assuming
that there is no atom at 0---that is, $\mu$ is supported on
$(0,\infty)$. Finally, it follows from the assumptions
that $\mu$ is an integrable measure with mean~$S_0$.\vadjust{\goodbreak}

Our goal is to now to use the knowledge of the call prices to find a
lower bound on the price of an option which has payoff
\[
F\biggl( \int_0^T \sigma_t^2 \dt\biggr) = F(
\langle \ln S\rangle_T ).
\]
Consider the discounted stock price,
\[
X_t = e^{-\int_0^t r_s \,\dif s} S_t = B_t^{-1} S_t.
\]
Under Assumption~\ref{assPrice}, $X_t$ satisfies the SDE
\[
\mathrm{d}X_t = X_t \sigma_t \,\dif W_t.
\]
Defining a time change $\tau_t = \int_0^t \sigma_s^2 \,\dif s$, and
writing $A_t$ for the right-continuous inverse, so that $\tau_{A_t} =
t$, we note that $\wti{W}_t = \int_0^{A_t} \sigma_s \,\dif W_s$ is a
Brownian motion with respect to the filtration $\wti{\F}_t =
\Fc_{A_t}$, and if we set $\wti{X}_t = X_{A_t}$, we have
\[
\mathrm{d}\wti{X}_t = \wti{X}_t \,\dif\wti{W}_t.
\]
In particular, $\wti{X}_t$ is now of a form where we may apply our
earlier results, using the target distribution arising from
\eqref{eqBL}, and noting also that $\wti{X}_0 = S_0$ and
$\wti{X}_{\tau_T} = X_T = B_T^{-1} S_T$.

We now define functions as in Section~\ref{secOptimality}, so that
$f(t) = F_+'(t)$ and \eqref{eqMdefn}--\eqref{eqHdefn} hold. Our aim
is to use \eqref{eqGFineq}, which now reads
%
%
\begin{equation} \label{eqGHineq}
\qquad G(X_{A_t},t) + H(X_{A_t}) = G(\wti{X}_t,t) + H(\wti{X}_t) \le
F(t) = F\biggl(\int_0^{A_t} \sigma_s^2 \,\dif s\biggr),
\end{equation}
to construct a sub-replicating portfolio. We shall first show that we
can construct a trading strategy that sub-replicates the
$G(\wti{X}_t,t)$ portion of the portfolio. Then we argue that we are
able, using a portfolio of calls, puts, cash and the underlying, to
replicate the payoff $H(X_T)$.\vspace*{1.5pt}

Since $G(\wti{X}_t,t)$ is a submartingale, we do not expect to be
able to replicate this in a completely self-financing manner. However,
by the Doob--Meyer decomposition theorem, and the martingale
representation theorem, we can certainly find some process
$\wti{\phi}_t$ such that
\[
G(\wti{X}_t,t) \ge G(\wti{X}_0,0) + \int_0^t \wti{\phi}_s \,\dif
\wti{X}_s
\]
and such that there is equality at $t = \tau_D$. Moreover, since
$G(\wti{X}_{\tau_D \wedge t}, \tau_D \wedge t)$ is a martingale, and
$G$ is $C^{2,1}$ in $D$, we have
\[
G(\wti{X}_{\tau_D \wedge t}, \tau_D \wedge t) = G(\wti{X}_0,0) +
\int_0^{\tau_D \wedge t} \frac{\partial G}{\partial x}(\wti
{X}_{\tau_D \wedge s},
\tau_D \wedge s) \,\dif\wti{X}_s.
\]
More generally, we would not expect $\frac{\partial G}{\partial x}$ to
exist everywhere in
$D^\complement$; however, if, for example,\vspace*{1pt} left and right derivatives\vadjust{\goodbreak}
exist, then
we could choose $\wti{\phi}_t \in[ \frac{\partial G}{\partial
x}(x-,t),
\frac{\partial G}{\partial x}(x+,t)]$ as our holding of the
risky asset (or
alternatively, but less explicitly, take $\wti{\phi}_t
= \partial/\partial x
[\E^{x,t}G(\wti{X}_{t+\delta},t_0+\delta)]$, for $t
\in
[t_0,t_0+\delta]$).

It follows that we can identify a process $\wti{\phi}_t$ with
\[
G(\wti{X}_{\tau_t},\tau_t) \ge G(\wti{X}_0,0) + \int_0^{\tau_t}
\wti{\phi}_s \,\dif\wti{X}_s
= G(X_0,0) + \int_0^t \wti{\phi}_{\tau_s} \,\dif X_s,
\]
where we have used, for example, Revuz and Yor~\cite{RevuzYor99}, Proposition~V.1.4. Finally, writing $\phi_s = \wti{\phi}_{\tau_s}$,
we have
\[
G(X_{t},\tau_{t}) \ge G(X_0,0) + \int_0^t \phi_s \,\dif X_s = G(X_0,0) +
\int_0^t \phi_s \,\dif(B_s^{-1} S_s).
\]
If we consider the self-financing portfolio which consists of holding
$\phi_s B_T^{-1}$ units of the risky asset, and an initial investment
of $G(X_0,0)B^{-1}_T-\phi_0 S_0 B_{T}^{-1}$ in the risk-free asset,
this has value $V_t$ at time $t$, where
\[
\mathrm{d}( B_t^{-1}V_t) = B_{T}^{-1}\phi_t
\,\dif(B_t^{-1}S_t),
\]
and therefore
\[
V_T = B_T \biggl( V_0 B_0^{-1} + \int_0^T B_{T}^{-1}\phi_s
\,\dif(B_s^{-1} S_s) \biggr) = G(X_0,0) + \int_0^T \phi_s
\,\dif X_s.
\]

We now turn to the $H(X_T)$ component in \eqref{eqGHineq}. If $H(x)$
can be written as the difference of two convex functions (so, in
particular, $H''(\mathrm{d}K)$ is a well-defined signed measure), we can
write
\begin{eqnarray*}
H(x) & =& H(S_0) + H_+'(S_0)(x-S_0) +
\int_{(S_0,\infty)}
(x-K)_+ H''(\mathrm{d}K) \\
& &{} + \int_{(0,S_0]} (K-x)_+ H''(\mathrm{d}K).
\end{eqnarray*}
Taking $x = X_T = B_T^{-1} S_T$, we get
\begin{eqnarray*}
H(X_T) &=& H(S_0) + H_+'(S_0)(B_T^{-1} S_T- S_0) + B_T^{-1}
\int_{(S_0,\infty)} (S_T - B_T K)_+ H''(\mathrm{d}K) \\
&& {} + B_T^{-1} \int_{(0,S_0]} (B_T K -S_T)_+ H''(\mathrm{d}K).
\end{eqnarray*}
This implies that the payoff $H(X_T)$ can be replicated at time $T$ by
``holding'' a portfolio of
%
%
\begin{eqnarray} \label{eqHport}
&& B_T^{-1}\bigl( H(S_0)- H_+'(S_0)S_0 \bigr) \mbox{ in cash;}\nonumber\\
&& B_T^{-1} H_+'(S_0) \mbox{ units of the asset;}
\nonumber
\\[-8pt]
\\[-8pt]
\nonumber
&& B_T^{-1} H''(\mathrm{d}K) \mbox{ units of the call with strike }
B_T K
\mbox{ for } K \in(S_0,\infty);\\
&& B_T^{-1} H''(\mathrm{d}K) \mbox{ units of the put with strike }
B_T K
\mbox{ for } K \in(0,S_0];\nonumber
\end{eqnarray}
where the final two terms should be interpreted appropriately. In
practice, the function $H(\cdot)$ can typically be approximated by a
piecewise linear function, where the ``kinks'' in the function
correspond to traded strikes of calls or puts, in which case the
number of units of each option to hold is determined by the change in
the gradient at the relevant strike. The initial cost of setting up
such a portfolio is well defined, provided
%
%
\begin{equation}
\label{eqHddcond}
\int_{(0,S_0]} P(B_T K) |H''|(\mathrm{d}K) +
\int_{(S_0,\infty)} C(B_T K) |H''|(\mathrm{d}K) < \infty,
\end{equation}
where $|H''|(\mathrm{d}K)$ is the total variation of the signed measure
$H''(\mathrm{d}K)$. We therefore shall make the following assumption:
\begin{assumption} \label{assHass}
The payoff $H(X_T)$ can be replicated using a suitable portfolio of
call and put options, cash and the underlying, with a finite price
at time~0.
\end{assumption}

We can therefore combine these to get the following theorem:
\begin{thmm} \label{thmHedge} Suppose that
Assumptions~\ref{assPrice},~\ref{assCalls} and~\ref{assHass}
hold, and suppose $F(\cdot)$ is a convex, increasing function with
$F(0) = 0$ and right derivative $f(t) = F_+'(t)$ which is
bounded. Then there exists an arbitrage if the price of an option
with payoff $F(\langle\ln S\rangle_T)$ is less than
%
%
\begin{eqnarray}
\label{eqFlb}
&& B_T^{-1} G(S_0,0) + B_T^{-1}H(S_0)  + B_T^{-1}\int_{(S_0,\infty)}
C(B_T K) H''(\mathrm{d}K)
\nonumber
\\[-8pt]
\\[-8pt]
\nonumber
&&\qquad {} + B_T^{-1}\int_{(0,S_0]} P(B_T K) H''(\mathrm{d}K),
\end{eqnarray}
where the functions $G$ and $H$ are as defined in \eqref{eqGdefn}
and \eqref{eqHdefn}, and are determined by the solution $\tau_D$ to
{SEP}$(\sigma,\delta_{S_0},\mu)$ for $\sigma(x) = x$, and where
$\mu$ is determined by~\eqref{eqBL}.

Moreover, this bound is optimal in the sense that there exists a
model which is free of arbitrage, under which the bound can be
attained.
\end{thmm}

\begin{pf}
It follows from Theorem~\ref{thmmainGBM} that, given $\mu$, we can
find a domain~$D$ and corresponding stopping time $\tau_D$ which
solves {SEP}$(\sigma,\delta_{S_0},\mu)$. Applying
Proposition~\ref{propGFineq} (and bearing in mind
Remark~\ref{rmkbndsigisid}), we conclude that the strategy
described above will indeed sub-replicate, and we can therefore
produce an arbitrage by purchasing the option, and selling short the
portfolio of calls, puts and the underlying given in
\eqref{eqHport}, and in addition, holding the dynamic portfolio
with $-\phi_tB_T^{-1}$ units of the underlying at time $t$. It is
not hard to check, given that~$f$ is bounded\vspace*{1pt} (and choosing the lower
limits in \eqref{eqZdefn} to be $S_0$ rather than $0$) that
$(Z'(\wti{X}_s) \sigma(\wti{X}_s))^2 \le(\wti{X}_s/\wti{X}_0 -
1)^2$, and hence that \eqref{eqqvbddcond} holds. Condition
\eqref{eqMassumption} also clearly holds. As a consequence, we do
indeed have a subhedge.\vadjust{\goodbreak}

To see that this is the best possible bound, we need to show that
there is a model which satisfies Assumption~\ref{assPrice}, has law
$\mu$ under $\Q$ at time $T$, and such that the subhedge is actually
a hedge. But consider the stopping time $\tau_D$ for the process~$\wti{X}_t$. Define the process
\[
X_t = \wti{X}_{{t}/{(T-t)} \wedge\tau_D}\qquad \mbox{for } t
\in[0,T]
\]
which corresponds to the choice of $\sigma_s^2 =
\frac{T-s+1}{(T-t)^2}\mathbf{1}_{\{{s}/{(T-s)} < \tau_D\}}$.
Since $\tau_D
< \infty$ a.s., then $X_T = \wti{X}_{\tau_D}$, $\tau_T = \tau_D$
and $S_t = X_{t}B_t$ is a price process satisfying
Assumption~\ref{assPrice} with
\[
F\biggl(\int_0^T \sigma_t^2 \,\dif t\biggr) = F(\tau_D).
\]
Finally, it follows from \eqref{eqGFequal} that at time $T$, the
value of the hedging portfolio exactly equals the payoff of the option.
\end{pf}

\begin{rmk}
The above results are given in the context of an increasing, convex
function, but there is also a similar result concerning increasing,
concave functions which can be derived. Consider a bounded,
increasing function $f$ as before, and define the function
\[
L(t) = \int_0^t \bigl( f(\infty) - f(s) \bigr) \,\dif s = f(\infty)t
- F(t).
\]
Using Theorem~\ref{thmHedge} and \eqref{eqloghedge}, it is easy to
see that the price of a contract with payoff $L(\langle \ln
S\rangle_T)$ must be bounded above by
\begin{eqnarray*}
&& 2 f(\infty)Q -  2 f(\infty) B_T^{-1} \log(S_0) - B_T^{-1}
G(S_0,0) - B_T^{-1}H(S_0) \\
&&\qquad {} - B_T^{-1}\int_{(S_0,\infty)} C(B_T K) H''(\mathrm{d}K)
- B_T^{-1}\int_{(0,S_0]} P(B_T K) H''(\mathrm{d}K),
\end{eqnarray*}
where $Q$ is the price of a log-contract [i.e., an option with
payoff $\ln(S_T)$]. As before, this upper bound is the best
possible, under a similar set of assumptions.
\end{rmk}

\begin{rmk}
An analogous result can be shown for \textit{forward start}
options. Suppose that the option has payoff
\[
F\biggl( \int_{S}^T \sigma_t^2 \dt\biggr) = F(
\langle S\rangle_T - \langle S\rangle_S )
\]
for fixed times $0 < S < T$. Then we can use the previous results
for general starting distributions to deduce a similar result to
Theorem~\ref{thmHedge} for forward start options, provided we
assume that there are calls traded at both $S$ and $T$. We use
essentially the same idea as above: we aim to hold a portfolio which
\mbox{(sub-)replicates} $G(X_t,\tau_t)$ for $t \in[S,T]$, and hold
the payoff $H(X_T)$ as a portfolio of calls. However, we now have
$\tau_t = \int_S^t \sigma_s^2 \,\dif s$, and so $\wti{X}_t = X_{A_t}$,
gives $\wti{X}_0 = X_S$ (recall that $A_t$ was assumed
right-continuous). The procedure is much as above, except that we
need to use the solution to Theorem~\ref{thmoptimality} with a
general target distribution, and the amount $G(\wti{X}_0,0)$ will be
a $\Fc_S$-random variable. The initial distribution $\nu$ can be
derived using the Breeden--Litzenberger formula \eqref{eqBL} at time
$S$. To ensure that we hold the amount $G(\wti{X}_0,0)$ at time $S$,
we observe that $G(\wti{X}_0,0) = G(X_S,0)$. Hence if, for example,
$G(x,0)$ can be written as the difference of two convex functions,
we can replicate this amount by holding a portfolio of calls and
puts with maturity $S$ in a similar manner to \eqref{eqHport}. The
remaining details follow as in the hedge described in
Theorem~\ref{thmHedge}
\end{rmk}

\begin{rmk} \label{rkcorvar}
We can also consider modifications to the realized
variance. Consider a slightly different time-change: suppose we set
\[
\tau_t = \int_0^t \sigma_s^2 \lambda(X_s) \,\dif s,
\]
for some ``nice'' function $\lambda(x)$, which in particular we
suppose is bounded above and below by positive constants. Then
following the computations above, we see that
\[
\wti{X}_t = X_{A_t} = \int_0^{A_t} X_s \lambda(X_s)^{-1/2}
( \sigma_s \lambda(X_s)^{1/2} \,\dif W_{s}) = \int_0^t
X_{A_s} \lambda(X_{A_s})^{-1/2} \,\dif\wti{W}_{s},
\]
and therefore $\mathrm{d}\wti{X}_t = \sigma(\wti{X}_t) \,\dif\wti{W}_t$,
where $\sigma(x) = x \lambda(x)^{-1/2}$. We then conjecture that it
is possible to extend Theorem~\ref{thmmainGBM} to cover this new
class of functions $\sigma(x)$ (the conditions that should be
imposed on $\lambda$ such that this result may be extended remains
an interesting question for future research). It would then be
possible to modify the above arguments to provide robust hedges on
convex payoffs of the form
\[
F\biggl(\int_0^T \sigma_s^2 \lambda(X_s) \,\dif s\biggr).
\]
An interesting special case of this would then be to give robust
bounds on the price of an option on corridor variance
%
%
\begin{equation} \label{eqcorvar}
F\biggl(\int_0^T \sigma_s^2 \mathbf{1}_{\{S_s \in[a,b]\}} \,\dif
s\biggr),
\end{equation}
by considering $\lambda(x) = \mathbf{1}_{\{x \in[a,b]\}}$, however this
would only work in the case where there are no discount rates (i.e.,
$B_t = 1$). In general, we can only give a tight lower bound for
options on
\[
F\biggl(\int_0^T \sigma_s^2 \mathbf{1}_{\{X_s \in[\wti{a},\wti
{b}]\}}
\,\dif s\biggr),
\]
although this does provide a lower bound for \eqref{eqcorvar} by
considering the case where $\wti{a} = a$ and $\wti{b} = B_T b$.
\end{rmk}

\section{Conclusions}

We conclude by summarizing the results, and describing some
interesting questions for future work. In this paper, we have given a
variational inequality representation of Root's solution to the
Skorokhod embedding problem, and provided a novel proof of optimality,
which allows us to construct a model-independent subhedge for options
on variance. We believe that our results provide interesting insights
into all three aspects of the work: the construction of solutions to
the Skorokhod embedding problem, proving optimality results for the
same and finally the connections with model-independent
hedging.

We also believe that there are interesting lines of research that now
arise. The construction opens up a number of questions regarding
Root's solution to the Skorokhod embedding problem: for example, what
can be said about the shape of the boundary? Under what conditions on
$\mu$ will the boundary be smooth? When does $R(x) \to0$ as $x \to
\pm\infty$? When is $R(x)$ bounded? Properties of free boundaries are
well studied in the analytic literature, and may be useful in
answering these questions. The connection to minimality and
noncentered target distributions raised in Remark~\ref{rknoncentred},
and the question asked at the end of this remark would also be
interesting lines for research.

The connection with optimal stopping noted in Remark~\ref{rmkoptstop}
is interesting, and obtaining a deeper understanding between optimal
stopping problems and optimal Skorokhod embeddings seems to be an
interesting area of research.

Another natural question concerns the upper bound/super-hedging
strategy. It has been remarked by Ob{\l}{\'o}j~\cite{Obloj04} and
Carr and Lee~\cite{CarrLee10} that a related construction of Rost
should provide a
suitable upper bound, but similar questions to those answered here
remain (although we hope to be able to provide some answers in
subsequent work). We note, however, that numerical evidence
(see Carr and Lee~\cite{CarrLee10}) seems to
suggest that the Root bounds may be more appropriate in the financial
applications. It would also be of interest to see to what extent these
model-independent bounds may be useful in practice. In
Cox and Ob{\l}{\'o}j~\cite{CoxObloj11}, an analysis of the use of
model-independent bounds
as a hedging strategy for barrier options was performed. A similar
analysis of the strategies derived in this work would also be of
interest.

Other questions that arise from the practical standpoint include how
to incorporate additional market information (e.g., calls at an
intermediate time~\cite{BrownHobsonRogers01b}), and how to adjust
for the fact that there will generally only be a finite set of quoted
calls; see~\cite{DavisOblojRaval10} for a related
question. Remark~\ref{rkcorvar} also suggests open questions
regarding more general choices of $\sigma(x)$.

\section*{Acknowledgment}
We are grateful to Sam Howison for a helpful discussion which has much
improved the material in Sections~\ref{secobs} and~\ref{secvarineq}.





%


\printaddresses

\end{document}